\documentclass[5p,10pt]{elsarticle}

\usepackage{amssymb}
\usepackage{amsmath}

\usepackage{xcolor}
\usepackage{multirow}
\usepackage{subcaption}
\usepackage{hyperref}

\journal{Computerized Medical Imaging and Graphics}

\begin{document}

\begin{frontmatter}



\title{Guidelines for Cerebrovascular Segmentation: Managing Imperfect Annotations in the context of Semi-Supervised Learning}


\author[CRESTIC,CREATIS]{Pierre~Roug\'e}
\author[IMT]{Pierre-Henri~Conze}
\author[CRESTIC]{Nicolas~Passat}
\author[CREATIS]{Odyss\'ee~Merveille}


\affiliation[CRESTIC]{organization={Université de Reims Champagne Ardenne, CReSTIC, Reims, France}}
            
\affiliation[CREATIS]{organization={Univ Lyon, INSA-Lyon, Universite Claude Bernard Lyon 1, CREATIS, Lyon, France}}

\affiliation[IMT]{
organization={IMT Atlantique, LaTIM UMR 1101, Inserm, Brest, France}
}

\begin{abstract}

Segmentation in medical imaging is an essential and often preliminary task in the image processing chain, driving numerous efforts towards the design of robust segmentation algorithms.
Supervised learning methods achieve excellent performances when fed with a sufficient amount of labeled data.
However, such labels are typically highly time-consuming, error-prone and expensive to produce.
Alternatively, semi-supervised learning approaches leverage both labeled and unlabeled data, and are very useful when only a small fraction of the dataset is labeled.
They are particularly useful for cerebrovascular segmentation, given that labeling a single volume requires several hours for an expert.
%
In addition to the challenge posed by insufficient annotations, there are concerns regarding annotation consistency. 
The task of annotating the cerebrovascular tree is inherently ambiguous. Due to the discrete nature of images, the borders and extremities of vessels are often unclear. Consequently, annotations heavily rely on the expert subjectivity and on the underlying clinical objective. 
%
These discrepancies significantly increase the complexity of the segmentation task for the model and consequently impair the results.
Consequently, it becomes imperative to provide clinicians with precise guidelines to improve the annotation process and construct more uniform datasets.
In this article, we investigate the data dependency of deep learning methods within the context of imperfect data and semi-supervised learning, for cerebrovascular segmentation. 
%
%
%
Specifically, this study compares various state-of-the-art semi-supervised methods based on unsupervised regularization and evaluates their performance in diverse quantity and quality data scenarios. Based on these experiments, we provide guidelines for the annotation and training of cerebrovascular segmentation models.

\end{abstract}




\end{frontmatter}


\section{Introduction}
\label{sec:Introduction}


In medical imaging, many pipelines used in clinical applications start by the segmentation of specific target structures.
Consequently, the segmentation task has been widely investigated in the image processing community. 
In particular, deep-learning methods have successfully alleviated performance bottlenecks and yield state-of-the-art results \cite{conze2023current}.
Currently, most segmentation approaches developed in this context relies on the U-Net architecture comprising convolutional \cite{ronneberger2015u, isensee2021nnu, azad2022medical} and/or transformers-based \cite{chen2021transunet, hatamizadeh2021swin, shamshad2023transformers} layers.

In the context of vascular segmentation \cite{DBLP:journals/mia/LesageABF09,DBLP:journals/cmpb/MocciaMHM18}, cerebrovascular segmentation stands out as one of the most intricate tasks. This complexity arises from the intricate nature of the brain's vascular network, which comprises numerous tortuous vessels with a complex structure, featuring cycles and a tree-like topology. While conventional U-Net models have been applied to cerebrovascular segmentation \cite{sanchesa2019cerebrovascular, livne2019u}, recent advancements have introduced more sophisticated architectures \cite{Valderrama2023, lin2023high, mou2021cs2, zhang2020cerebrovascular}. However, these approaches rely on supervised learning which requires a large quantity of labeled data, specifically the annotation of the whole brain vascular network across numerous 3D images.

Unfortunately, these annotations have to be produced and/or carefully reviewed by medical experts which is heavily time-consuming.
As a result, most cerebrovascular segmentation datasets contain a limited number of labeled training data, thereby limiting the performance of trained models.



In addition to the lack of labeled data, another significant challenge arises from the considerable variability in cerebrovascular labels. This variability comes from several factors. Firstly, the definition of a cerebrovascular label varies depending on the specific application of interest \cite{ciecholewski2021computational}. Some datasets concentrate solely on major vessels, such as the circle of Willis (TopCow \cite{yang2023benchmarking}), while others include all arteries but exclude veins (Bullitt \cite{aylward2002initialization}, IXI \cite{ixidataset}). Despite the adoption of a global annotation policy, the limit regarding where to stop annotating can be ambiguous and vary from one dataset to another. For instance, when stating ``all arteries are labeled'', some datasets may or may not include the ophthalmic arteries and the arteries supplying blood to the scalp.

Even though the same annotation policy is shared between two datasets, variations in the extent of labels may arise depending on the annotator. Firstly, due to the discrete nature of annotations, the status (inside or outside the vascular label) of voxels on the vessel border becomes ambiguous and is subject to the annotator's discretion. Secondly, a similar issue arises when labeling a vessel near the image resolution limit. The annotators may choose to stop labeling at different positions.

\begin{figure}
    \centering
    \includegraphics[width=\linewidth]{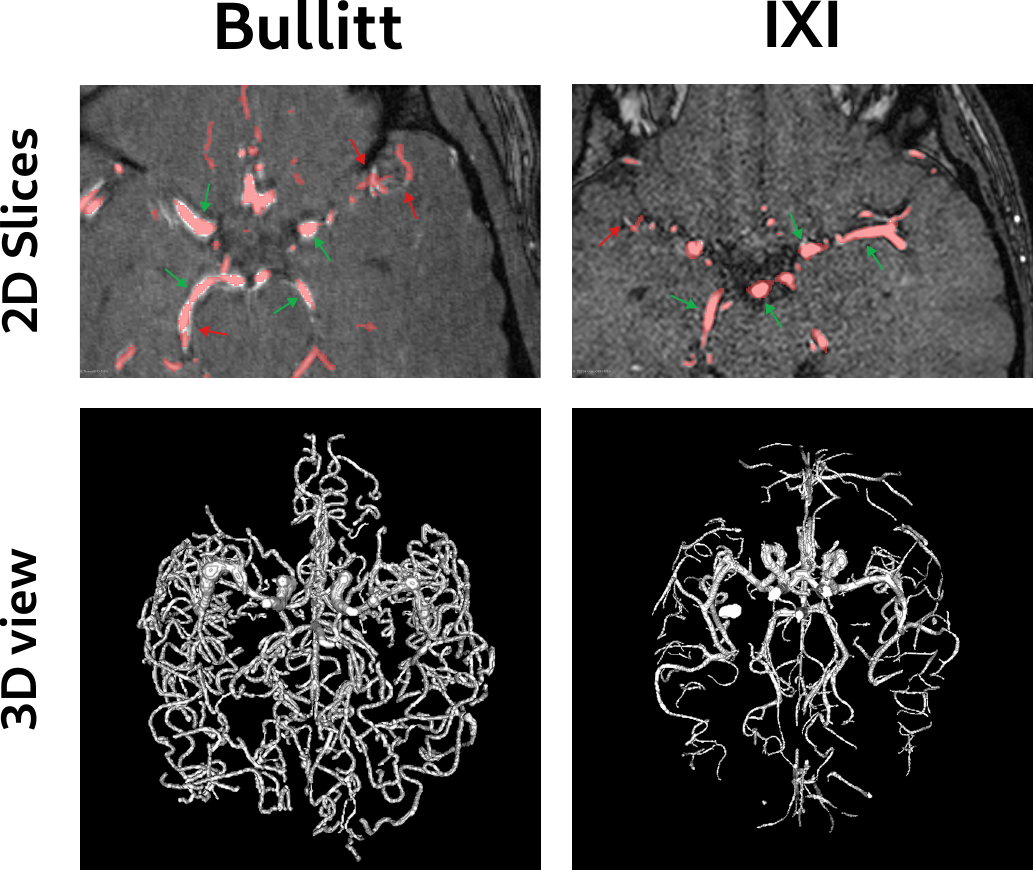}
    \caption{Illustration of noisy labels and concept shift problem arising on two cerebrovascular datasets : Bullitt \cite{aylward2002initialization} (left) and IXI \cite{ixidataset} (right). Top row : 2D slices with labels overlaid in light red, vessels with ambiguous boundaries indicated by green arrows and noisy labels (such as  missing vessels or vessel disconnections), marked by red arrows. Bottom row: 3D view showing the disparities in the extent of the labels for the same global annotation policy ``labeling all cerebrovascular arteries''.}
    \label{fig:label_shift}
\end{figure}

These numerous sources of vascular label variability create what is commonly referred to as concept shift \cite{liu2022deep, kouw2018introduction}. During training, the model tries to learn the concept of ``vessel'' based on the provided labels. However, this concept may undergo changes depending on the annotator responsible for its creation (inter-expert variability) or even the moment of the label creation (intra-expert variability). This concept shift is often even more important when using several datasets during the same training, as slight changes in the annotation policy may lead to systematic bias in the concept of vessel. For example, the extent of the artery border as well as the arteries included in the label in the Bullitt dataset compared to the IXI dataset are significantly different (Fig.~\ref{fig:label_shift}). 
This concept shift tends to reduce the performance of the segmentation model and the stability of the training. Additionally, it causes problems when evaluating a model on a dataset exhibiting a concept shift compared to the one used during training.

Secondly, extensively labeling cerebrovascular structures is a laborious task, making it prone to significant variations in the quality of annotations. Notably, common imperfections observed in cerebrovascular labels include missed vessels, disconnections or holes within a labeled vessel, and the inadvertent labeling of veins in datasets intended for arterial annotation (illustrated in Fig.\ref{fig:label_shift}). Another common imperfection is the absence of spatial continuity in labels, especially when annotations are exclusively conducted in a single 2D plane, typically the axial one.
These imperfections generate noisy labels, further complicating the training process of the segmentation model.


This problem of imperfect annotations is closely linked with the problem of annotations scarcity. When confronted with a limited quantity of labeled data, the absence of diversity in annotations can result in deep learning models exhibiting an overfitting behavior on the specific concept and/or noise present on the labels of the training dataset, potentially introducing the biases discussed above. Hence, there is a notable interest in examining these issues together.

As discussed before, the challenges of data scarcity, concept shift and noisy labels are more pronounced in cerebrovascular segmentation compared to other segmentation applications, thus they represent an important performance bottleneck for supervised segmentation methods. Yet, to the best of our knowledge, this has not been studied.



To address both the scarcity of cerebrovascular annotations and the challenges related to their quality and consistency, there has been a growing interest in semi-supervised methods.
These methods tackle the problem of lack of annotation by leveraging unlabeled data, alongside labeled data, to increase the model's performance. Three distinct strategies can be identified \cite{jiao2022learning}: (1) generating pseudo-labels to train a supervised model \cite{shi2021inconsistency, wang2022ssa, thompson2022pseudo}, (2) using unlabeled data to perform an unsupervised regularization, (3) employing unlabeled data to learn prior knowledge through self-supervised tasks \cite{you2022simcvd, zheng2019semi}. 
The most prevalent strategy among these is the unsupervised regularization approach, further categorized into four types \cite{jiao2022learning}: (1) consistency learning \cite{tarvainen2017mean, luo2021semi, yu2019uncertainty, wu2021semi, lei2022semi, bai2023bidirectional}, (2) co-training \cite{luo2022semi, xia20203d}, (3) adversarial learning \cite{li2020shape, zhang2017deep, hou2022semi}, and (4) entropy minimization \cite{hang2020local}. In this study, we only focus on the unsupervised regularization approaches, as they also address the challenge of data imperfection by limiting overfitting. 


Semi-supervised unsupervised regularization methods have been applied to cerebrovascular segmentation.
For instance, \cite{xie2022semi} proposed a mean-teacher framework where the consistency loss was guided by a region connectivity model to introduce anatomical constraints. Also, \cite{chen2022generative} proposed to use, in addition to the segmentation model, a generative model that reconstructs the input image from the segmentation outputs. A reconstruction loss was then computed to add another unsupervised regularization.


Employing semi-supervised cerebrovascular segmentation strategies seems indeed appealing to cope with the lack of labeled data. Indeed, this approach not only boost the model performance but also regularize the training and limit overfitting. Nonetheless, it remains imperative to include a sufficient amount of labeled data to enable the network to grasp the concept of a vessel. Moreover, this sufficient amount of labeled data highly depends on the quality of the annotations. 
Therefore, when encountering a new cerebrovascular segmentation challenge, finding a balance between the quantity of labeled data, their quality, and the chosen approach becomes paramount.

In this article, our goal is to formulate guidelines to assist the community in finding this balance when addressing a cerebrovascular segmentation problem. In particular, we try to address the following questions:

\begin{itemize}
    \item When is it more interesting to use semi-supervised approach than a fully supervised approach for cerebrovascular segmentation? 
    \item What type of semi-supervised approach should be used for cerebrovascular segmentation?
    \item Can semi-supervised learning methods effectively mitigate overfitting on the annotated data they use?
    \item Is the quality of annotation more crucial than quantity?
    \item What types of annotation imperfections most significantly degrade segmentation results and should be primarily avoided?
\end{itemize}

To do so, we identified five state-of-the-art semi-supervised segmentation approaches for medical images, along with a fully supervised baseline and compared them on two different cerebrovascular datasets. We specifically examine their behavior with respect to the number of available labeled data. We also study the sensibility of these approaches to the specific labeled samples that are used during training.
%
%
We finally investigate the impact of several common imperfection types found in cerebrovascular labels: missing vessels, over- and under-vessel border segmentation. 

The remainder of this article is organized as follows.
Sect.~\ref{sec:methods} presents the datasets and semi-supervised methods used in this study.
Sect.~\ref{sec:results} details both the experimental setup and the outcomes of the conducted experiments.
Finally, in Sect.~\ref{sec:discussion_conclusion}, we discuss the results and conclusions of this study.

\section{Materials and methods}
\label{sec:methods}

In this section, we present the datasets as well as the employed semi-supervised learning approaches. 

\begin{figure*}
    \centering
    \includegraphics[width=\linewidth]{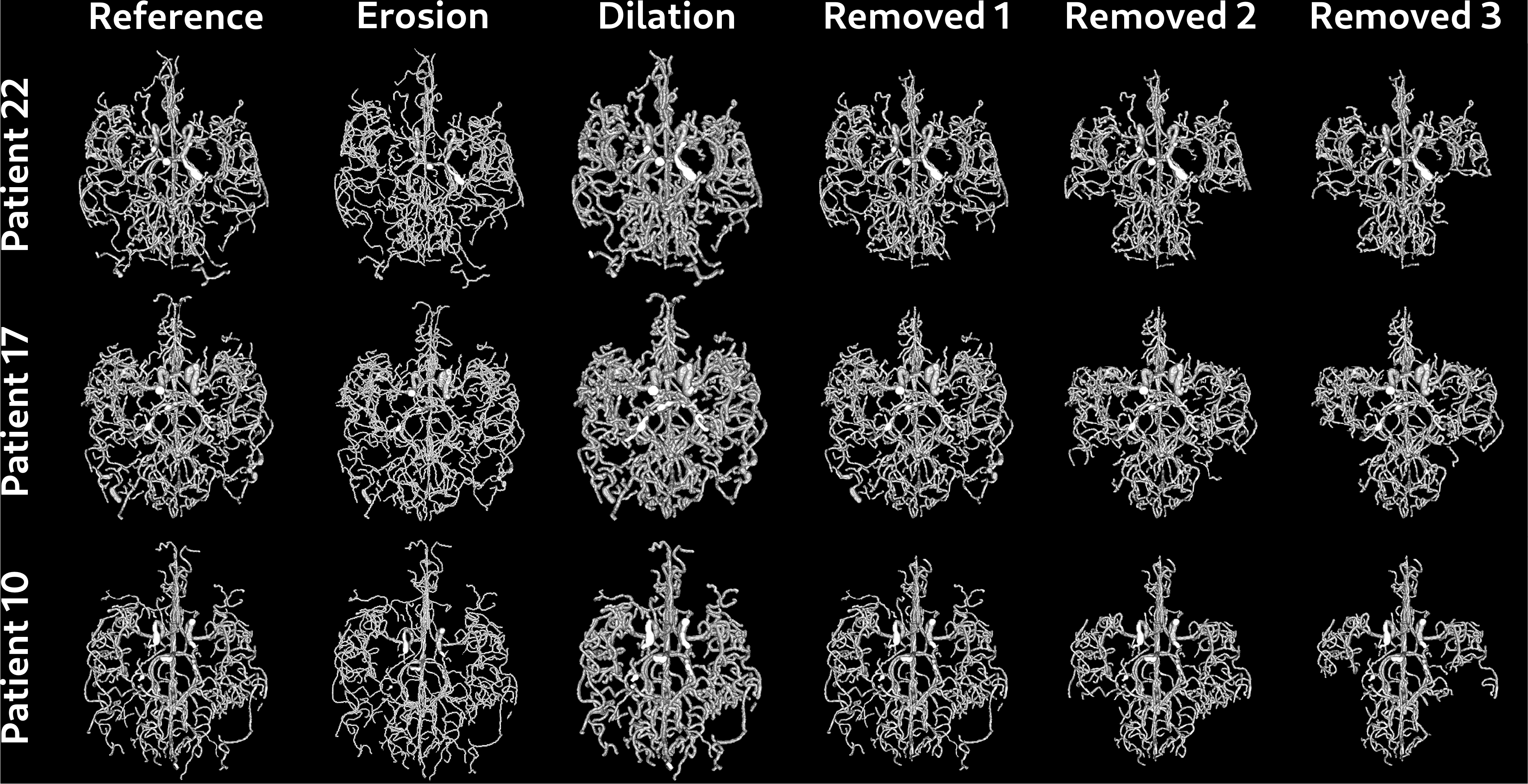}
    \caption{Visualization for three different patients of the different deteriorations carried out on the Bullitt dataset to investigate the impact of concept-shift in cerebrovascular segmentation.}
    \label{fig:deterioration}
\end{figure*}

\subsection{Datasets}

We used the public dataset Bullitt \cite{aylward2002initialization} composed of $109$ Time-of-Flight Magnetic Resonance Angiography (TOF-MRA) images and $34$ annotations.
All volumes have the same resolution  with voxels of size $0.51 \times  0.51 \times 0.80$ mm$^{3}$ and a shape of $448 \times 448 \times 128$.
We split this dataset into $94$ volumes (including $19$ annotated ones) for training, and $15$ annotated volumes for test.

We also used the public dataset IXI \cite{ixidataset} originally containing nearly $600$ TOF-MRA images acquired from three different centers.
We chose to only use the data from the Guy’s Hospital, London, UK.
This subset represents $316$ volumes with $15$ annotations.
All volumes have the same resolution with voxels  of size of $0.47 \times  0.47 \times 0.80$ mm$^{3}$ and a shape of $512 \times 512 \times 100$.
We split this dataset into $311$ volumes, including $10$ annotated volumes for training, and $5$ annotated volumes for test.

\subsection{Metrics}
\label{sec:metrics}

In this study, our focus was restricted to two metrics. We first used the Dice score ($DSC$) \cite{dice1945measures}, defined as:
\begin{equation}
    DSC = \frac{2 \cdot  tp}{2\cdot tp + fp + fn}
\end{equation}
where $tp$, $fp$, $fn$ are the true positives, false positives and false negatives, respectively.

Additionally, we incorporated the clDice score \cite{shit2021cldice} that was built upon the Dice paradigm to mitigate bias towards large vessels, thus offering a more adapted evaluation for vessel segmentation, especially in the case of complete vascular networks that exhibit varying radius branches.

To compute the clDice score, we first extract the centerlines $C_{GT}$ and $C_{P}$ from the ground truth segmentation mask $S_{GT}$ and the predicted segmentation $S_{P}$, respectively. Then, we compute the ratio of $C_{P}$ inside $S_{GT}$, called \textit{topology precision} ($Tp$), and the ratio of $C_{GT}$ inside $S_{P}$, called \textit{topology sensitivity} ($Ts$):
\begin{align}
    Tp &= \frac{|C_{P} \cap S_{GT}|}{C_{P}}\\
    Ts &= \frac{|C_{GT} \cap S_{P}|}{C_{GT}}
    \label{eq:tprec_tsens}
\end{align}
Then, clDice is defined as the harmonic mean of $Tp$ and $Ts$:
\begin{equation}
        clDice =
        \frac{2 \cdot Tp \cdot Ts}{Tp + Ts}
\end{equation}

\subsection{Methods}

In this section, we present the five semi-supervised learning methods which were selected for comparison purposes. These methods all belong to the group of unsupervised regularization methods.
They differ in their regularization strategy: consistency upon input noise \cite{tarvainen2017mean, yu2019uncertainty}, adversarial consistency \cite{li2020shape}, task level consistency \cite{luo2021semi}, and mutual consistency \cite{wu2021semi}.

\subsubsection{Mean-teacher}

\begin{figure}
    \centering
    \includegraphics[width=\linewidth]{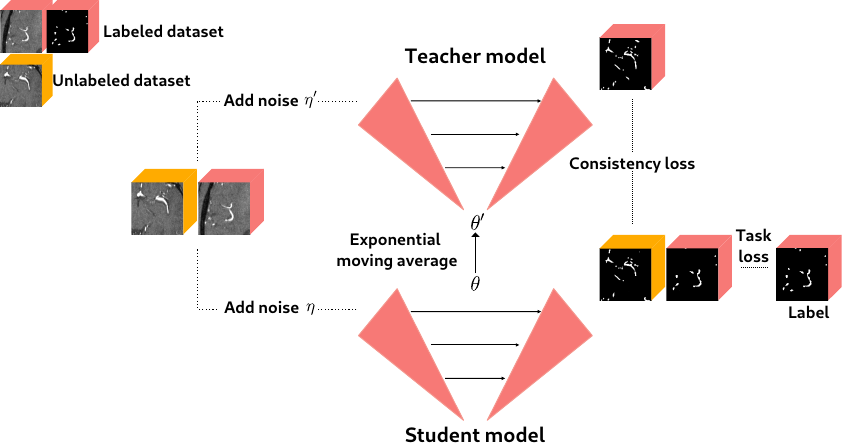}
    \caption{Illustration of the mean-teacher model leveraging both labeled and unlabeled data for cerebrovascular segmentation.}
    \label{fig:mean_teacher}
\end{figure}

Among the semi-supervised methods, the mean-teacher model \cite{tarvainen2017mean} is a popular backbone (Fig.~\ref{fig:mean_teacher}).
In this framework, two different models are used, namely a teacher and a student.
Only the student model is trained through backpropagation, while the teacher weights are updated by the exponential moving average of the student weights.
The training procedure leverages the unlabeled data by performing an unsupervised regularization through a consistency loss.
More precisely, the input data is disturbed with two different random Gaussian noises, leading to two slightly different inputs for the student and teacher model, respectively.
Then, the consistency loss is used between the outputs of the student and teacher models, forcing these outputs to be consistent despite the perturbation.
This regularization allows to leverage unlabeled data and improve the model performance.
The model loss $\mathcal{L}$ to optimize is:
\begin{equation}
    \mathcal{L} = \mathcal{L}_{s} + \lambda_{c} \cdot \mathcal{L}_{c}
\end{equation}
with $\mathcal{L}_{s}$ the supervised segmentation loss computed on the labeled samples, $\mathcal{L}_{c}$ the consistency loss computed on the unlabeled samples and $\lambda_{c} \geq 0$ a parameter setting the weight of the consistency loss. For every method, the parameter $\lambda_{c}$ evolves during training and follows a Gaussian warming up function.

\begin{figure*}[t]
    \centering
    \includegraphics[width=\linewidth]{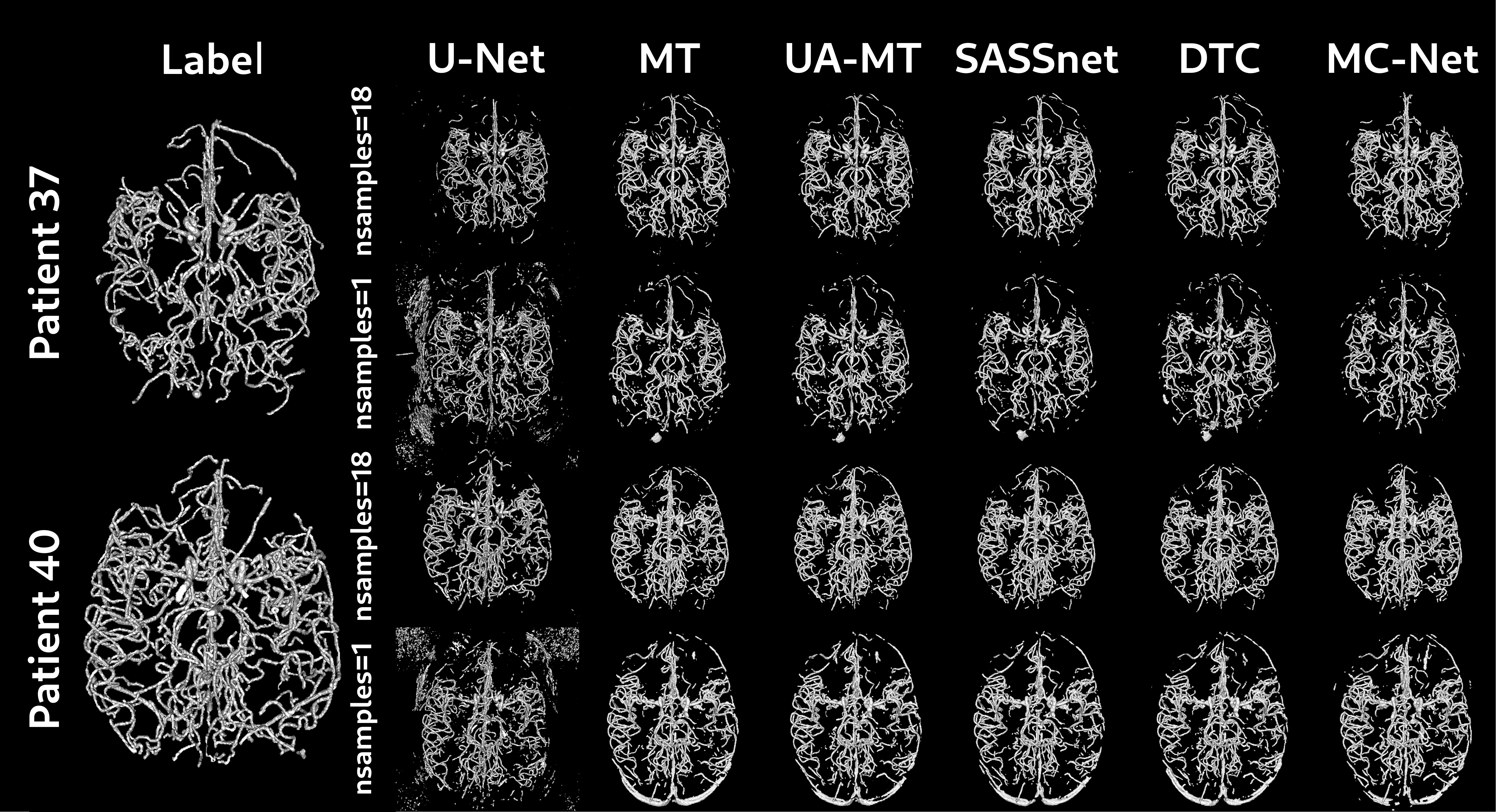}
    \caption{Visual results for Bullitt dataset from experiment 1. Segmentation results are presented for two patients, each method and two dataset compositions.}
    \label{fig:visual_results_xp1_Bullitt}
\end{figure*}

\subsubsection{Uncertainty-aware mean-teacher}

The uncertainty-aware mean-teacher \cite{yu2019uncertainty} relies on the mean-teacher framework.
It proposes to monitor the consistency regularization by using uncertainty.
The uncertainty on the output of the teacher model is estimated with Monte-Carlo dropout.
The model performs $N$ forward passes with different dropouts, and the uncertainty is estimated by the entropy of these $N$ predictions.

More formally, by setting $P_{n}^{c}$ the probability volume of the $c$-th class of the $n$-th forward pass, the uncertainty volume is defined as:

\begin{equation}
    U = - \sum_{c}\mathcal{U}_{c}\log(\mathcal{U}_{c})
\end{equation}
with:
\begin{equation}
    \mathcal{U}_{c} = \frac{1}{N} \sum_{n}P_{n}^{c}
\end{equation}

This uncertainty is used to monitor the consistency regularization.
The computation of the consistency loss is restricted to voxels falling below a given threshold.
Throughout training, this threshold is dynamically adjusted, ensuring that the consistency loss is computed only on voxels where the teacher's prediction is reliable in the early stages of training, thereby enhancing regularization robustness.
Conversely, as training progresses, the consistency loss is expanded to incorporate a broader set of ambiguous voxels, thereby capturing more complex information.

Let us consider $\mathcal{P}_t$ the prediction of the teacher and  $\mathcal{P}_s$ the prediction of the student.
We mask them with the uncertainty volume, as follows:
\begin{align}•••••••••••••
    \mathcal{P}_t^{mask} = \mathbb{I}( U < \tau).\mathcal{P}_t \\
     \mathcal{P}_s^{mask} = \mathbb{I}( U < \tau).\mathcal{P}_s
\end{align}

with $\tau$ a dynamically adjusted threshold and $\mathbb{I}$ the indicator function.
The threshold $\tau$ is defined by a Gaussian warming up function:
\begin{equation*}
    \tau(t) = \ln(2)\left(\frac{3}{4} + \frac{1}{4}\exp\left(-5\left(1-\frac{t}{t_{max}}\right)^{2}\right)\right)
\end{equation*}
where $t$ is the current training step and $t_{max}$ the final training step.

The final model loss to optimize can be written as:
\begin{equation}
    \mathcal{L} = \mathcal{L}_{s} + \lambda_{c} \cdot \mathcal{L}_{c}(\mathcal{P}_t^{mask}, \mathcal{P}_s^{mask})
\end{equation}
with $\mathcal{L}_{s}$ the supervised segmentation loss computed on the labeled samples, $\mathcal{L}_{c}$ the consistency loss and $\lambda_{c} \geq 0$ a parameter setting the weight of the consistency loss.

\subsubsection{Shape-aware semi-supervised network}

The shape-aware semi-supervised (SASSnet) model \cite{li2020shape} uses adversarial learning to enforce consistency of unlabeled and labeled data predictions.
In this method, the regularization is not done on the segmentation output but on a signed distance map output to enforce geometric shape constraints. 

Towards this goal, a multi-task architecture is built with a shared encoder and a decoder with two outputs heads: one for segmentation and one for signed distance map regression. The supervised loss $\mathcal{L}_s$ is then the combination of the segmentation loss $\mathcal{L}_{seg}$ and the mean squared error loss for the signed distance map $\mathcal{L}_{sdm}$:
\begin{equation}
    \mathcal{L}_s = \mathcal{L}_{seg} + \alpha.\mathcal{L}_{sdm}
\end{equation}
with $\alpha \geq 0$ a hyperparameter balancing the two losses.

Then, a discriminator is designed to distinguish between the labeled and unlabeled datasets. To be more precise, this discriminator receives the original image and the signed distance map as inputs and is required to classify whether the input originates from labeled data or unlabeled data.
Optimizing this discriminator in an adversarial learning scheme helps to learn shape-aware features shared by both labeled and unlabeled datasets.

The final training objective to optimize is written as:
\begin{equation}
    \mathcal{L}(\theta, \zeta) = \mathcal{L}_{s}(\theta) + \lambda_{c} \cdot \mathcal{L}_{adv}(\theta, \zeta)
\end{equation}
with $\theta$ the parameters of the segmentation network, $\zeta$ the parameter of the discriminator, $\mathcal{L}_{adv}$ the adversarial loss and  $\lambda_{c} \geq 0$ a weight to control the importance of each loss.
The optimization process is performed following:
\begin{equation}
    \min_{\theta} \max_{\zeta} \mathcal{L}(\theta, \zeta)
\end{equation}

\begin{figure*}[t]
    \centering
    \includegraphics[width=\linewidth]{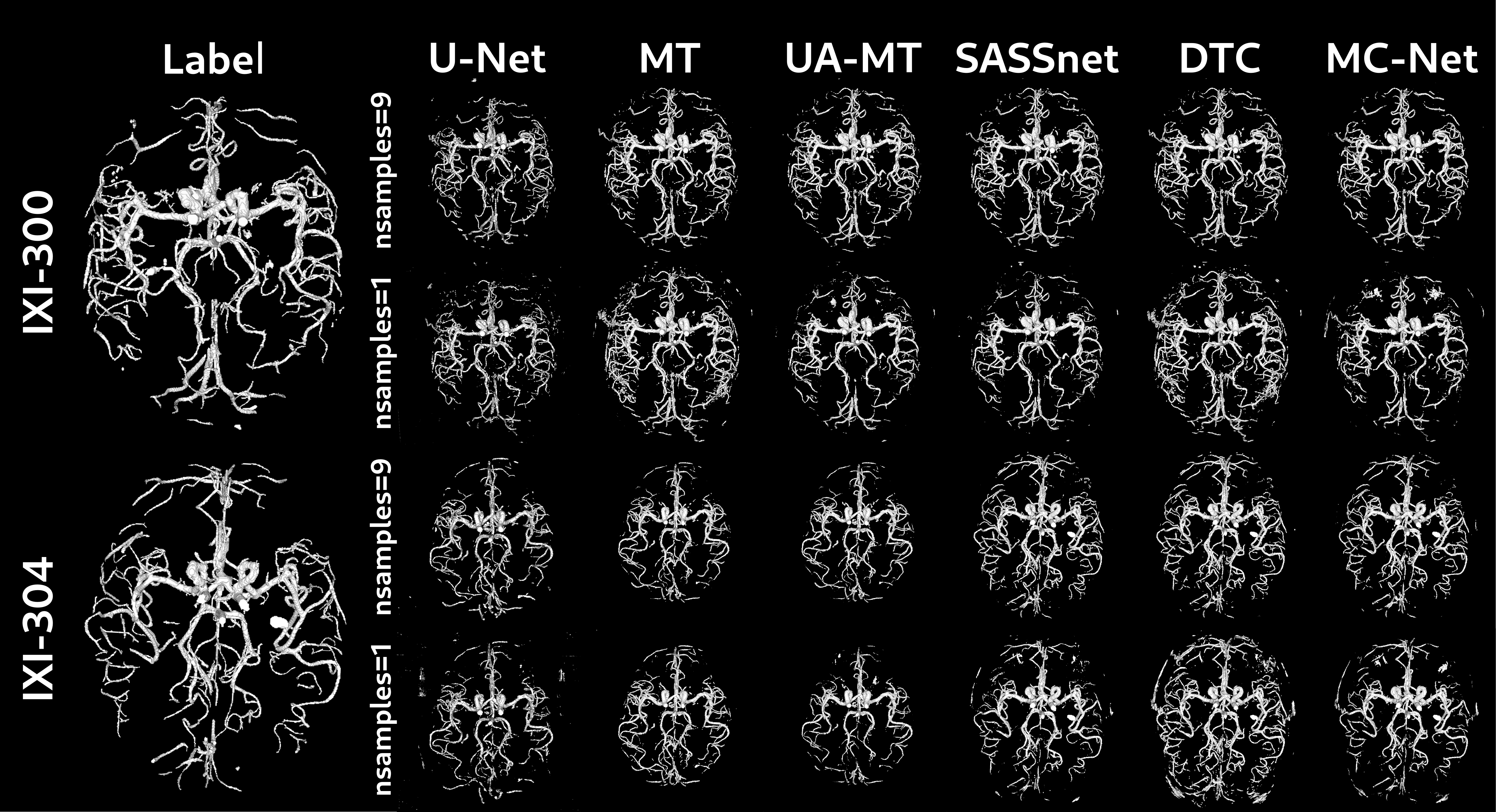}
     \caption{Visual results for IXI dataset from experiment 1. Segmentation results are presented for two patients, each method and two data dataset compositions.}
    \label{fig:visual_results_xp1_IXI}
\end{figure*}

\subsubsection{Dual-task consistency network}

The dual-task consistency (DTC) network is based on task level consistency.
Similarly to SASSnet \cite{li2020shape}, the framework consists of a voxel-wise segmentation head and a signed distance map regression head. The supervised loss is a combination of the segmentation loss $\mathcal{L}_{seg}$ and the mean squared error loss for the signed distance map $\mathcal{L}_{sdm}$:
\begin{equation}
    \mathcal{L}_s = \mathcal{L}_{seg} + \alpha \cdot \mathcal{L}_{sdm}
\end{equation}

However, in contrast to SASSnet, the unsupervised regularization is performed by consistency regularization. Indeed, the signed distance map is transformed into a segmentation map, leading to a slightly different result than the segmentation head output.
A consistency loss is finally used to enforce consistency between these two segmentation maps.

The mapping from the signed distance map to the segmentation map is approximated by the function $\mathcal{T}(z_{i})$ defined for a signed distance map value $z_i$ of pixel $i$ by:
\begin{equation}
    \mathcal{T}(z_i) = (1 + e^{-k.z_i})^{-1}
\end{equation}
with $k$ a parameter set to $-1500$.
Then, the consistency loss $\mathcal L_{c}$ is computed between the obtained segmentation map and the output of the segmentation head.
The final model loss is defined as:
\begin{equation}
    \mathcal{L}_{total} = \mathcal{L}_{s} + \lambda_{c} \cdot \mathcal{L}_{c}
\end{equation}
with $\mathcal{L}_{s}$ the segmentation loss computed on labeled data, $\mathcal{L}_{c}$ the consistency loss and $\lambda_{c} \geq 0$ a parameter setting the contribution of the consistency loss.

\subsubsection{Mutual consistency network}

The mutual consistency network (MC-Net) \cite{wu2021semi} consists of one shared encoder and two different decoders. The two decoders have a slightly different architecture. Since their outputs are different, the consistency loss enforces the consistency between them.
More precisely, if we consider $\mathcal{P_A}$ the prediction of the first decoder and $\mathcal{P_B}$ the prediction of the second decoder, $\mathcal{P_A}$ and $\mathcal{P_B}$ are transformed into soft labels $s\mathcal{P_A}$ and $s\mathcal{P_B}$ through a sharpening function:
\begin{equation}
    s\mathcal{P} = \frac{\mathcal{P}^\frac{1}{T}}{\mathcal{P}^\frac{1}{T} + (1 - \mathcal{P})^\frac{1}{T}}
\end{equation}
with $T$ a parameter to control the sharpening, set to $0.1$.

Then $s\mathcal{P_A}$ is used to supervise $\mathcal{P_B}$ while $s\mathcal{P_B}$ is used to supervise $\mathcal{P_A}$.
This enforces the prediction consistency between the two decoders and promote low-entropy predictions.
Finally, the loss to optimize is defined by: 
\begin{equation}
    \mathcal{L} = \mathcal{L}_{segA} + \mathcal{L}_{segB} + \lambda_{c} \cdot \big(\mathcal{L}_{c}(s\mathcal{P_A}, \mathcal{P_B}) + \mathcal{L}_{c}(s\mathcal{P_B}, \mathcal{P_A})\big)
\end{equation}
with $\mathcal{L}_{segA}$ the supervised segmentation loss for the first decoder, $\mathcal{L}_{segB}$ the supervised segmentation loss for the second decoder, $\mathcal{L}_{c}$ the consistency loss and $\lambda_{c} \geq 0$ a parameter setting the weight of the consistency loss.

\section{Experiments and results}
\label{sec:results}

In this section, we explain in details the experiments carried out to investigate the semi-supervised learning approaches described above and to provide annotations guidelines for  cerebrovascular segmentation.

\begin{figure}[t]
    \begin{subfigure}[b]{0.5\textwidth}
        \centering
        \includegraphics[width=\linewidth]{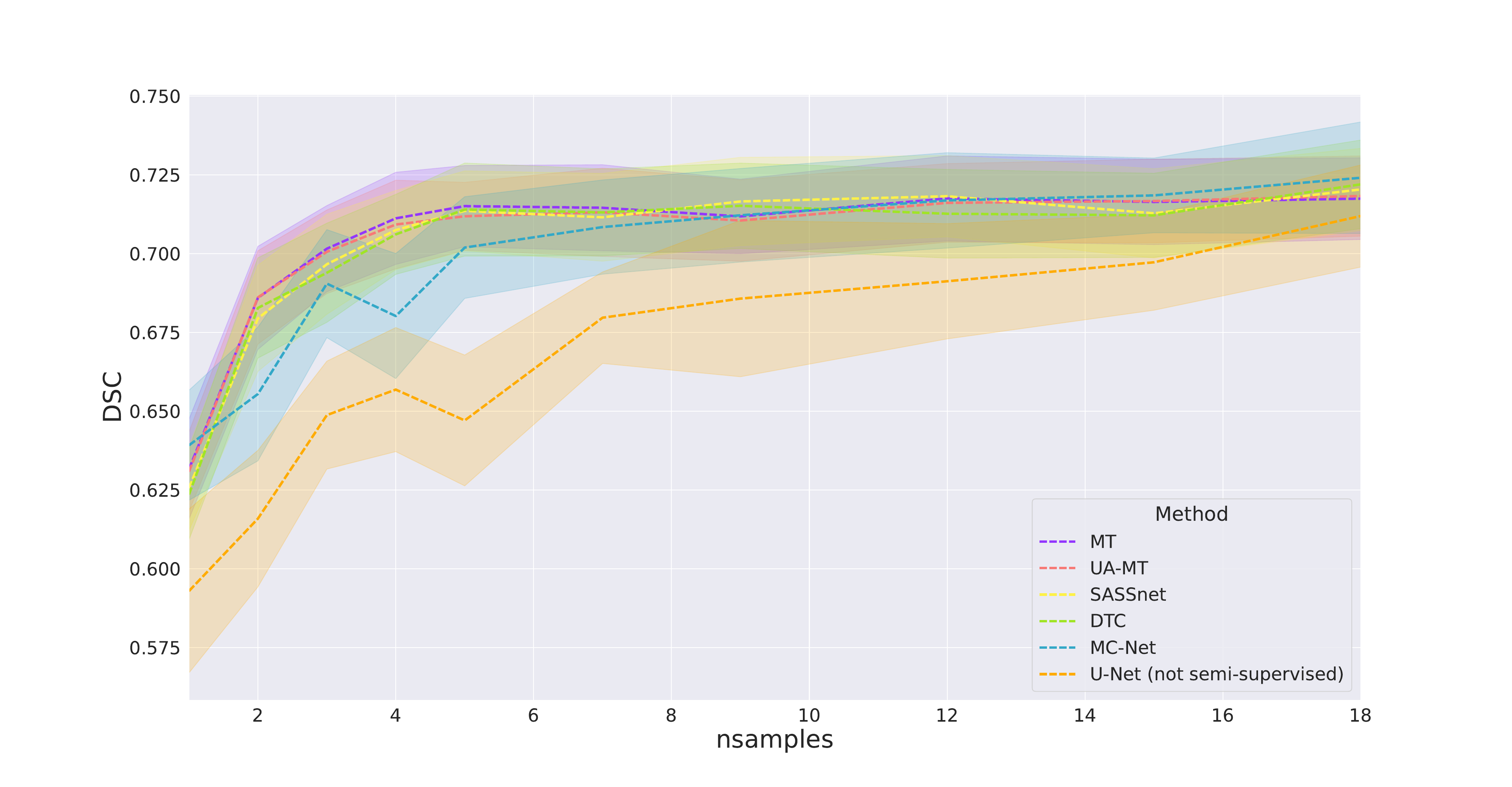}
        \caption{$DSC$}
    \end{subfigure}
    \begin{subfigure}[b]{0.5\textwidth}
        \centering
        \includegraphics[width=\linewidth]{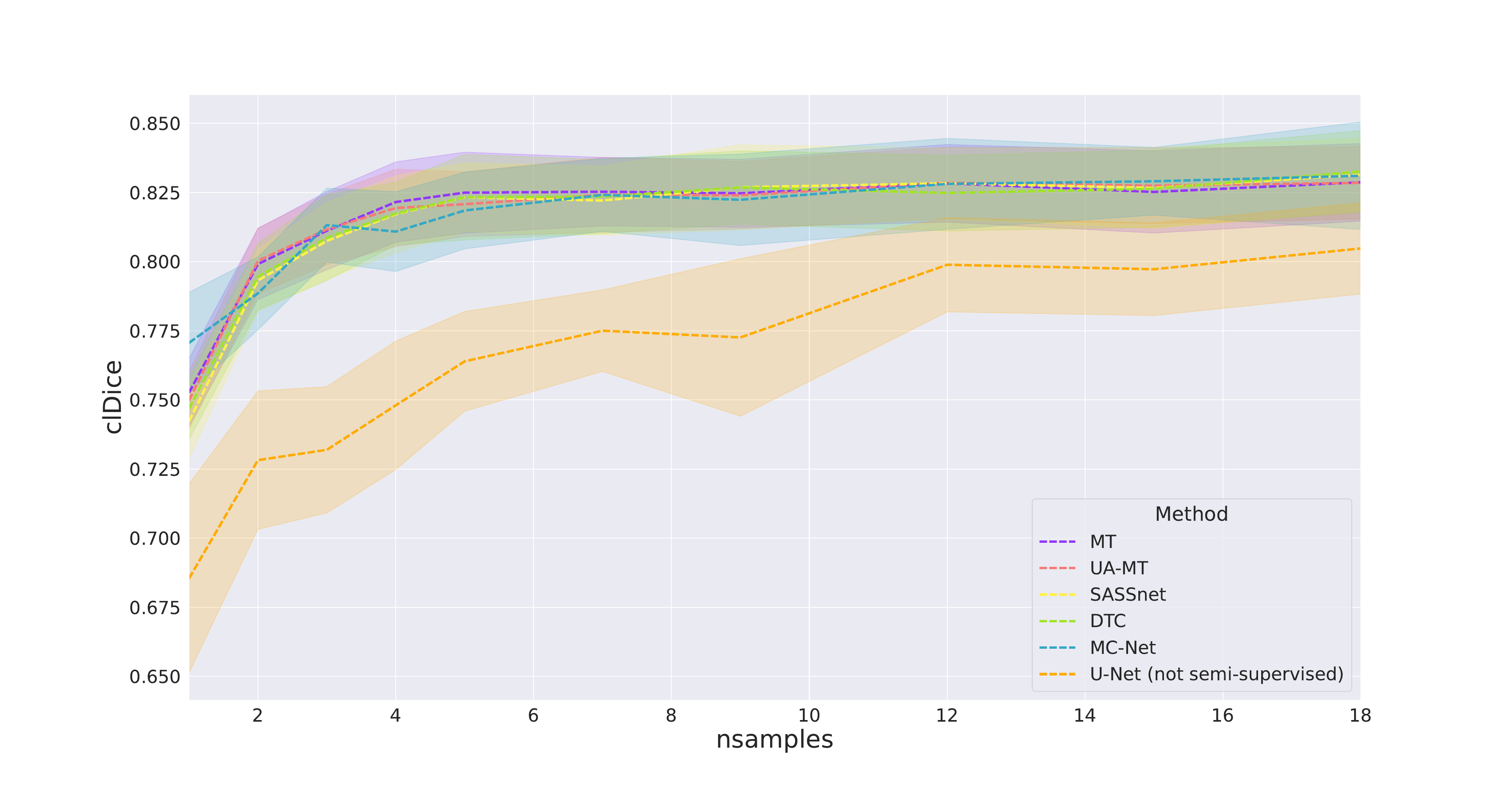}
        \caption{$clDice$}
    \end{subfigure}
    \caption{$DSC$ (a) and $clDice$ (b) scores obtained on the test set for the Bullitt dataset, depending on the dataset composition. Each curve corresponds to a different method. The transparent bands represent the standard deviation.}
    \label{fig:res_xp1_bullitt}
\end{figure}

\begin{figure}[t]
    \begin{subfigure}[b]{0.5\textwidth}
        \centering
        \includegraphics[width=\linewidth]{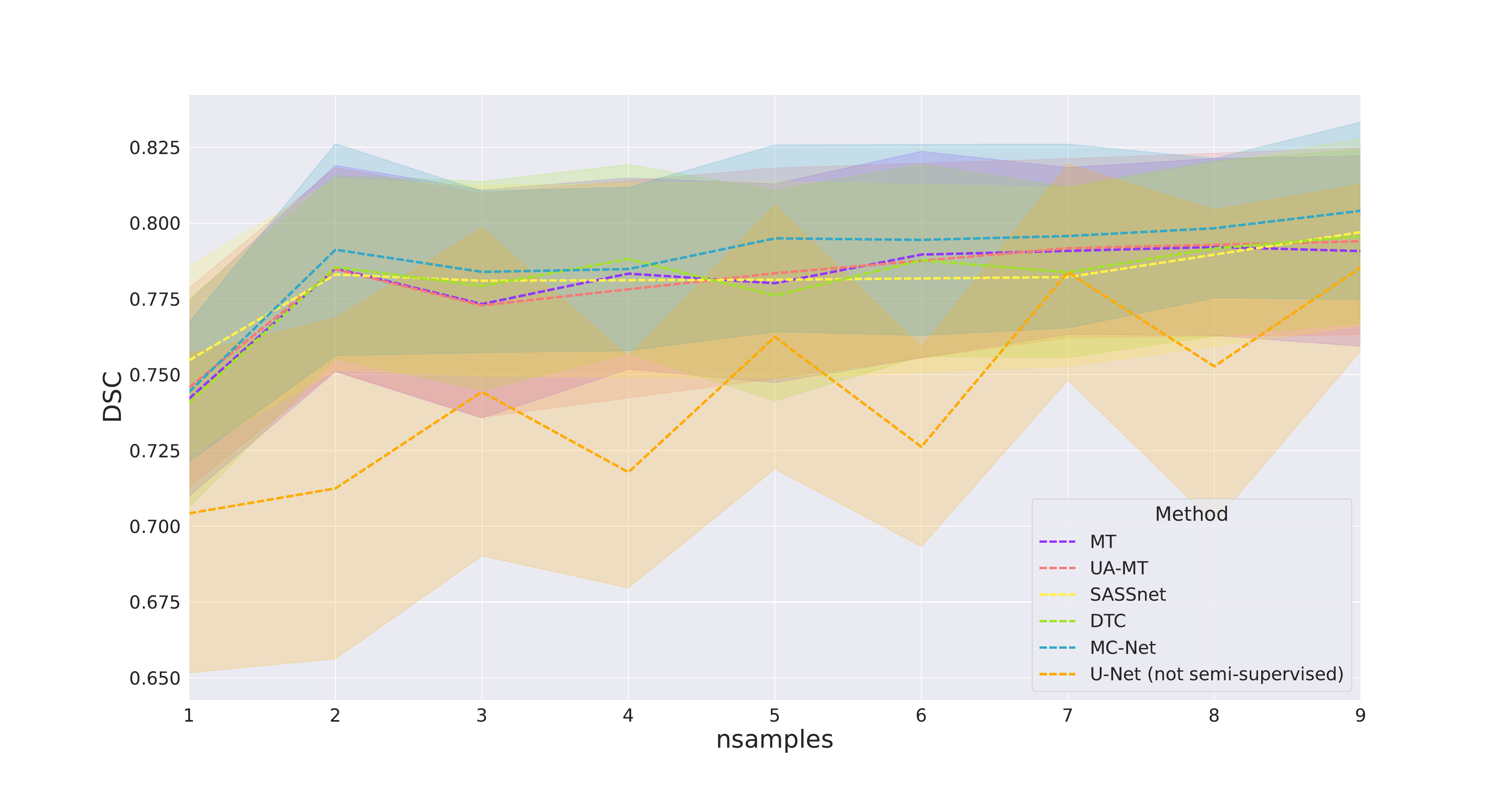}
        \caption{$DSC$}
    \end{subfigure}
    \begin{subfigure}[b]{0.5\textwidth}
        \centering
        \includegraphics[width=\linewidth]{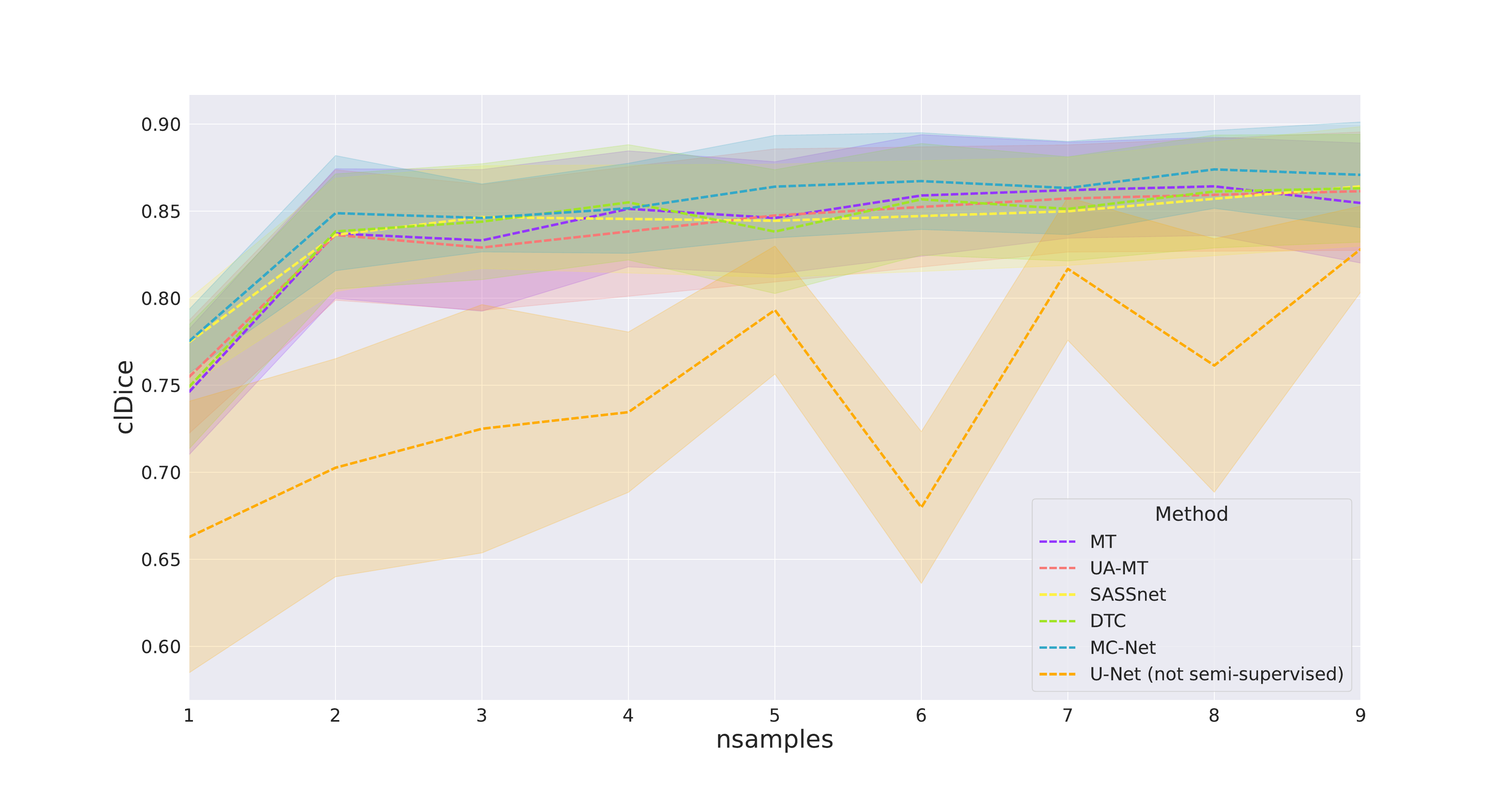}
        \caption{$clDice$}
    \end{subfigure}
    \caption{$DSC$ (a) and $clDice$ (b) scores obtained on the test set for the IXI dataset, depending on the dataset composition. Each curve corresponds to a different method. The transparent bands represent the standard deviation.}
    \label{fig:res_xp1_IXI}
\end{figure}

\subsection{Experimental setup}

We performed three sets of experiments, aiming at addressing the questions raised in the introduction.


\subsubsection{Experiment 1}

The purpose of the first experiment is to conduct a comparative analysis among various semi-supervised learning methods and assess their performance in comparison to a supervised baseline.
More specifically, we aim to clarify in which context (i.e. with how many data available) semi-supervised learning brings a significant improvement with respect to the supervised baseline.
To this end, we evaluated all the methods on two datasets and with several dataset compositions (Tab.~\ref{tab:firs_experience_setup}). Specifically, we increased the number of labeled data used for training while keeping the dataset size constant.

\begin{table}[!t]
    \centering
        \caption{Overview of the dataset sizes and the dataset compositions employed in Experiment 1.}
    \label{tab:firs_experience_setup}
    \begin{tabular}{lcc}
        \hline
        &  Bullitt & IXI\\
        \hline
        Total dataset size & $109$ & $316$\\
        Training set size & $94$ &  $311$\\
        Test set size & $15$ & $5$\\
        Labeled data in the training set & $19$ &  $10$\\
        \hline
        \multirow{10}{*}{Various dataset compositions} & (1, 93) & (1, 310)\\
        \multirow{10}{*}{(num labeled, num unlabeled)} & $(2, 92)$ & $(2, 309)$ \\
        & $(3, 91)$ & $(3, 308)$ \\
        & $(4, 90)$ & $(4, 307)$ \\
        & $(5, 89)$ & $(5, 306)$ \\
        & $(7, 87)$ & $(6, 305)$ \\
        & $(9, 85)$ & $(7, 304)$ \\
        & $(12, 82)$ & $(8, 303)$ \\
        & $(15, 79)$ & $(9, 302)$ \\
        & $(18, 76)$ &  - \\
        \hline
    \end{tabular}
\end{table}

\begin{figure}[t]
    \begin{subfigure}[b]{0.5\textwidth}
        \centering
        \includegraphics[width=\linewidth]{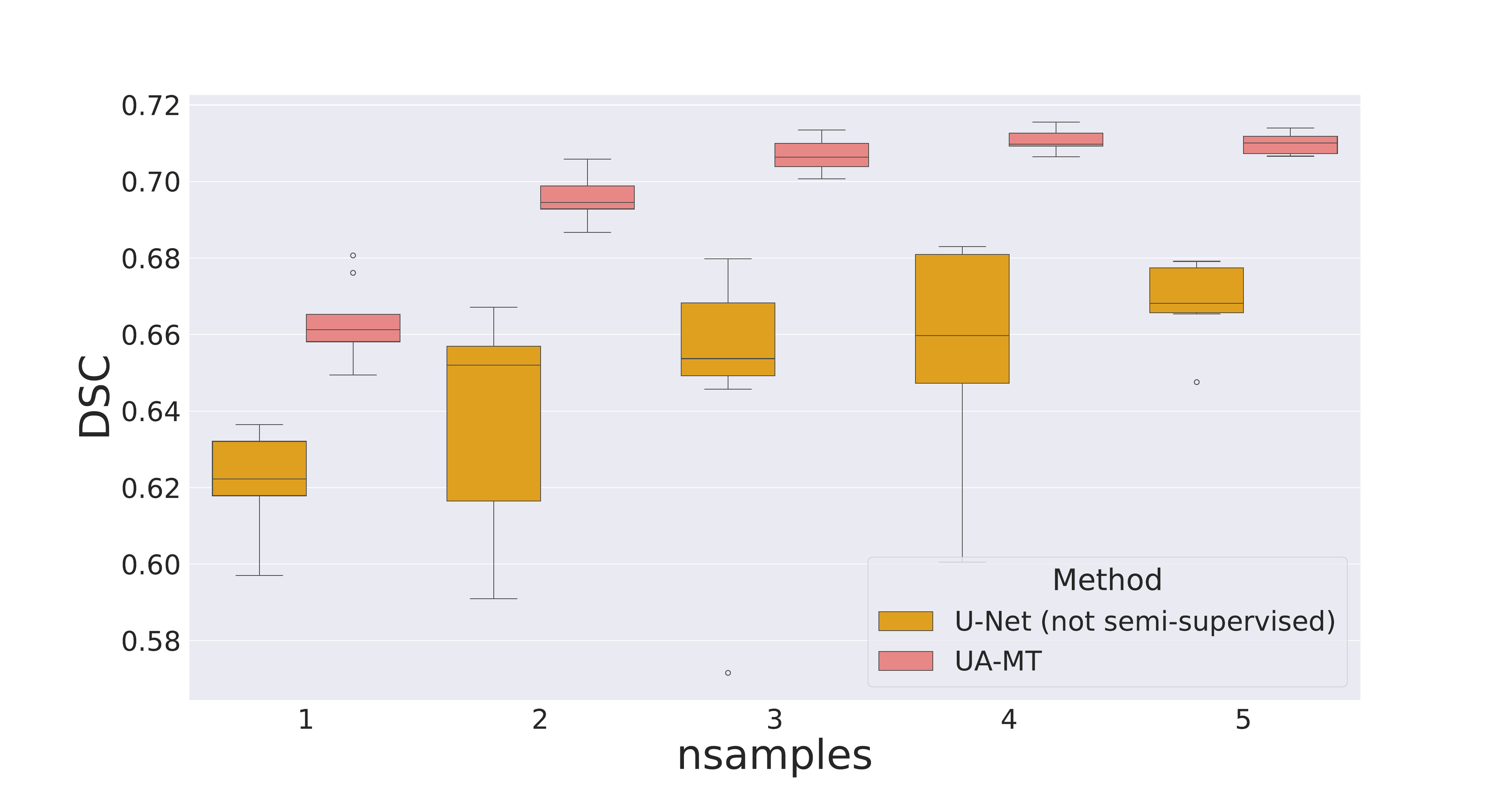}
        \caption{$DSC$}
    \end{subfigure}
     \begin{subfigure}[b]{0.5\textwidth}
        \centering
        \includegraphics[width=\linewidth]{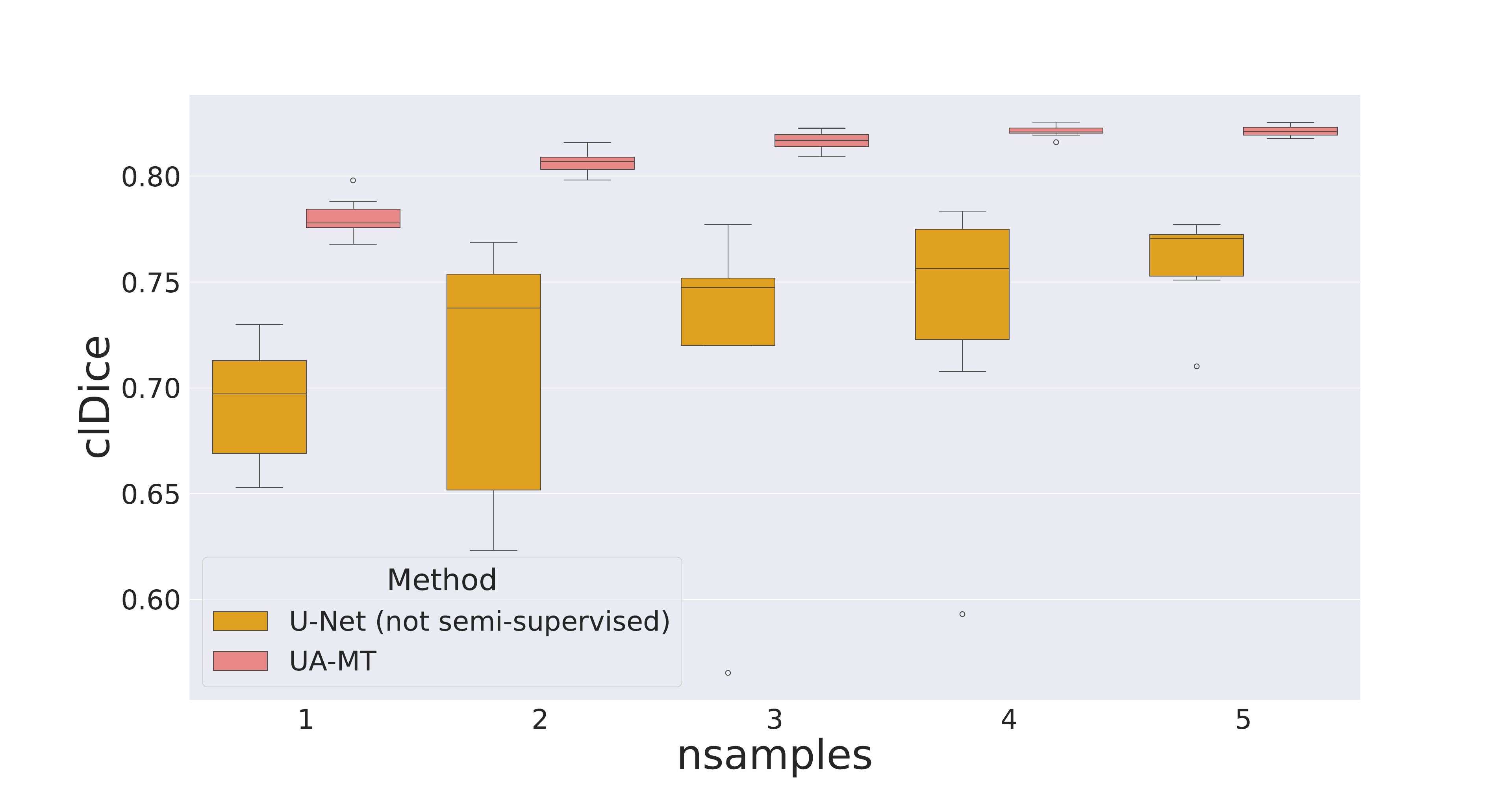}
        \caption{$clDice$}
    \end{subfigure}
    \caption{Boxplots of $DSC$ (a) and $clDice$ (b) calculated on the different training seeds for Bullitt. Each color represent a different method (UA-MT and U-Net).}
    \label{fig:seed_impact_Bullitt}
\end{figure}

\begin{figure}[t]
    \begin{subfigure}[b]{0.5\textwidth}
        \centering
        \includegraphics[width=\linewidth]{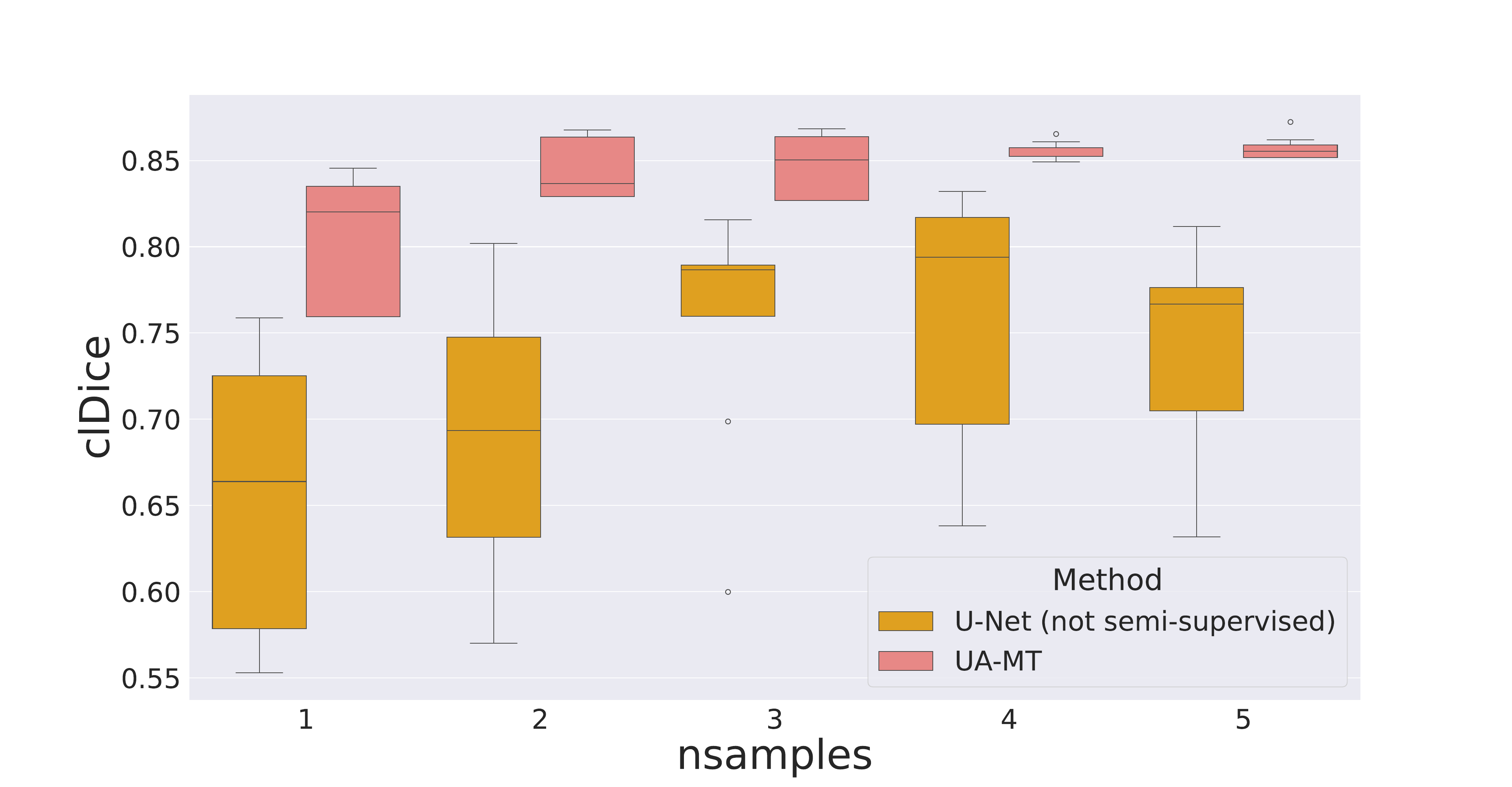}
        \caption{$DSC$}
    \end{subfigure}
    \begin{subfigure}[b]{0.5\textwidth}
        \includegraphics[width=\linewidth]{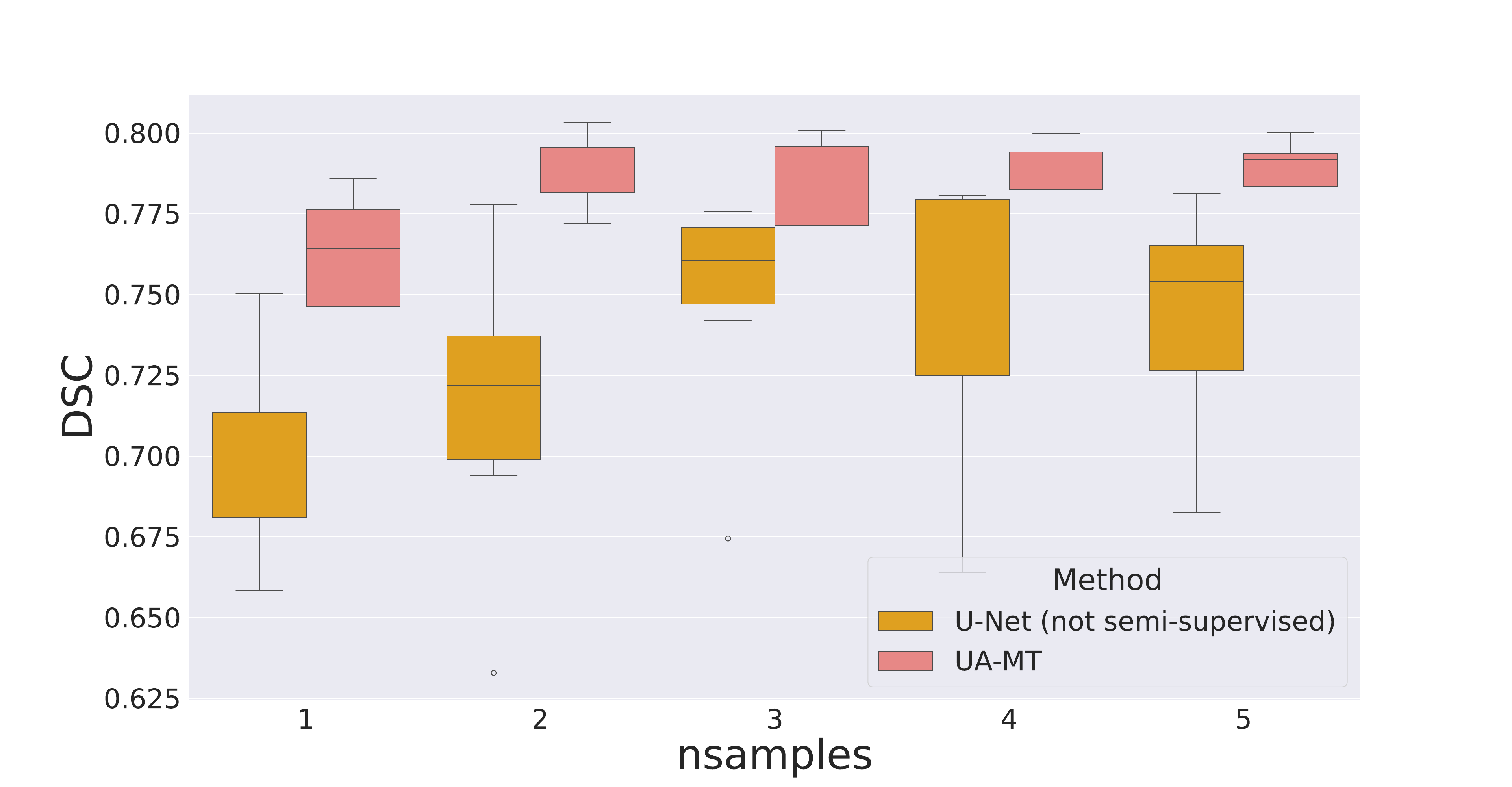}
        \caption{$clDice$}
    \end{subfigure}
    
    \caption{Boxplots of $DSC$ (a) and $clDice$ (b) calculated on the different training seeds for IXI. Each color represent a different method (UA-MT and U-Net).}
    \label{fig:seed_impact_IXI}
\end{figure}
\subsubsection{Experiment 2}

In a low-data regime, the risk of overfitting is very high and the performance of the model highly depends on the data used for training.
We aim at investigating the extent to which the semi-supervised learning framework succeed to reduce both overfitting and dependency to the data used for training. 

To achieve this goal, we compare a representative semi-supervised approach, the UA-MT method \cite{yu2019uncertainty}, to the supervised baseline with several different data seeds.
We choose the UA-MT method among the semi-supervised methods as it is the most established method in the literature.

In this experiment, we selected the first five data compositions with the least number of labeled data ($1$ to $5$). For each data composition and for each method (UA-MT and the baseline), we trained nine models, each with a different seed to select the labeled samples. Our goal is to quantify and compare the variability of both models, depending on the choice of the seed.

\subsubsection{Experiment 3}


In this experiment, our objective is to study the impact of both type and degree of label imperfection on the segmentation results. We aim to identify whether certain imperfections have a lesser impact on segmentation outcomes. If so, the annotation process and review can be tailored to prioritize key factors that significantly affect results, optimizing both the annotation process and model performance. Ultimately, our goal is to formulate annotation guidelines for enhanced efficiency and model performance.


We identified three main categories of annotations imperfections: over-estimated vessel radius, under-estimated vessel radius, and missing small vessels (or part of vessels).


We simulated each type of imperfection as follows:
\begin{itemize}
    \item Under-segmentation : all annotations are eroded with a ball of radius $1$ voxel and we made sure that it does not introduce any vessel disconnection. This scenario is referred to as ``Erosion".
    \item Over-segmentation: all annotations are dilated with a ball of radius $1$. This scenario is referred to as ``Dilation".
    \item Missing vessels:  we manually removed an increasing amount of vessels on the distal parts of the vascular network to create three levels of missing vessels. These scenarios are referred to as ``Removed 1, Removed 2, Removed 3".
\end{itemize}
These simulated imperfections were performed on $5$ volumes. We then trained the UA-MT model \cite{yu2019uncertainty}, on these deteriorated datasets for increasing number of available labeled samples and tested it on the original dataset.





\begin{figure}[t]
    \begin{subfigure}[b]{0.5\textwidth}
        \centering
        \includegraphics[width=\linewidth]{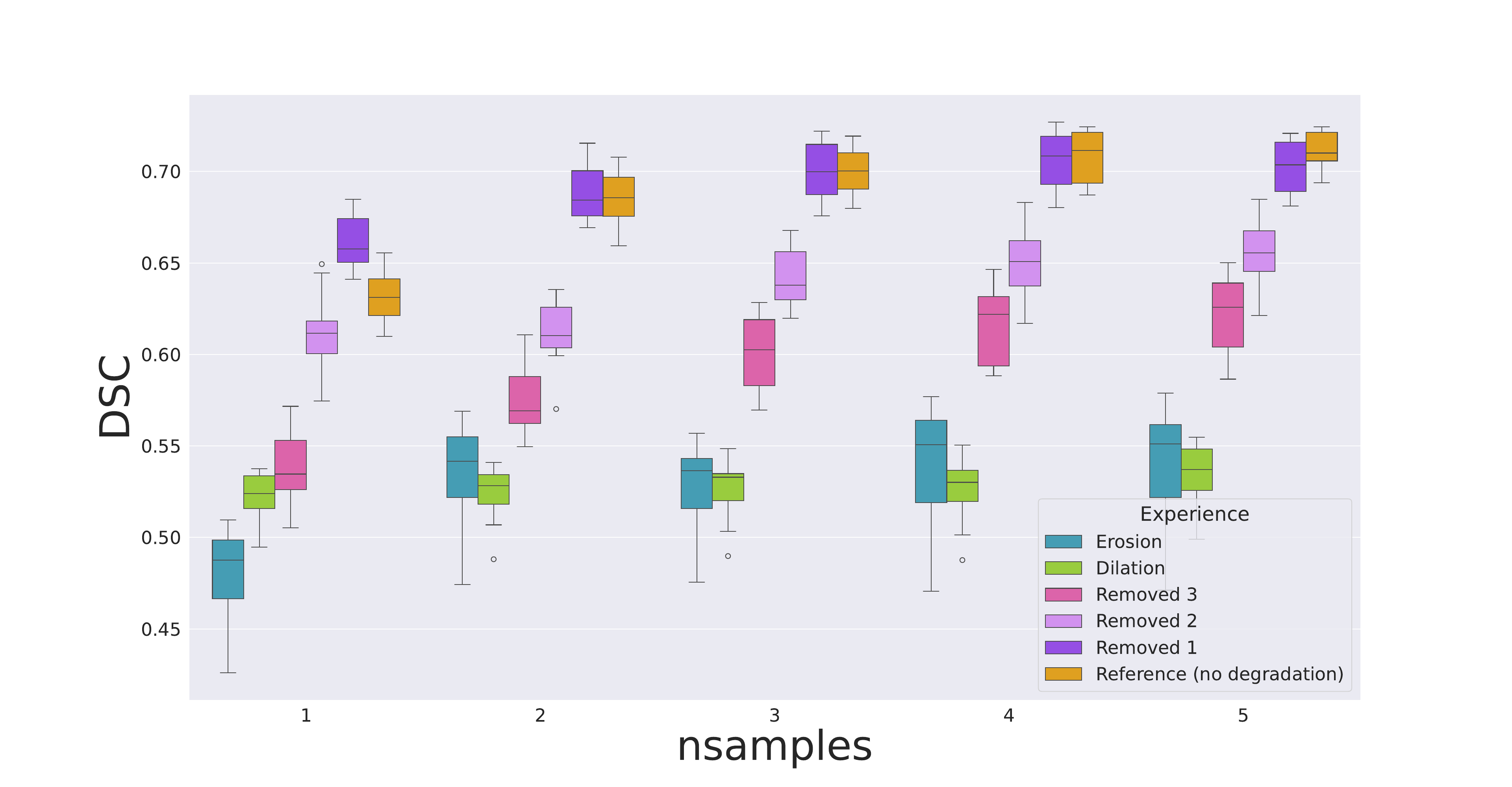}
        \caption{$DSC$}
    \end{subfigure}
    \begin{subfigure}[b]{0.5\textwidth}
        \centering
        \includegraphics[width=\linewidth]{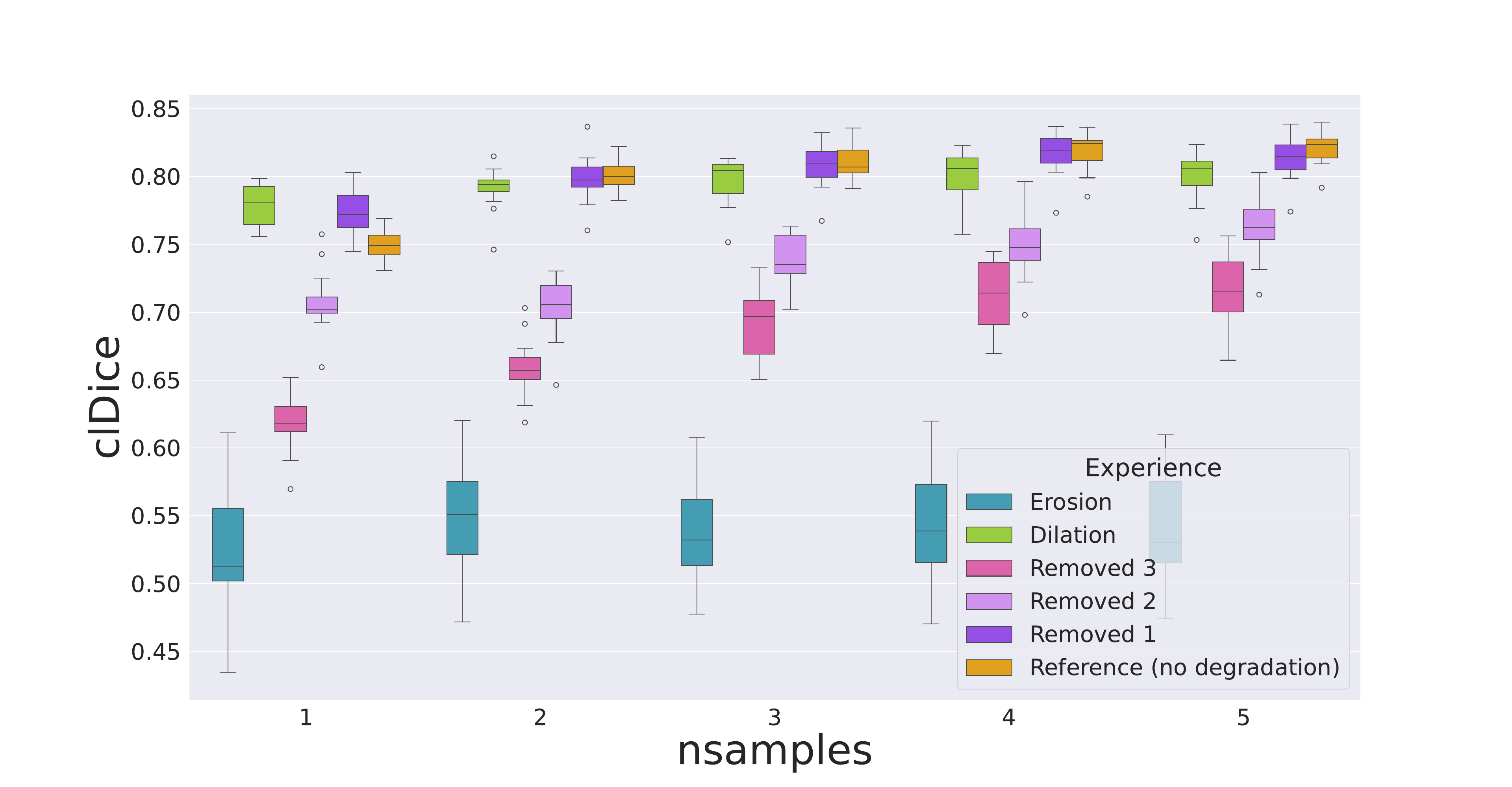}
        \caption{$clDice$}
    \end{subfigure}
    \caption{Boxplots of $DSC$ (a) and $clDice$ (b) on test set for every experiment, depending on the dataset composition. Each color correspond to a different experiment.}
    \label{fig:deterioration_results}
\end{figure}

\subsection{Implementation details}

To conduct all these experiments, we relied on the official implementation of the evaluated semi-supervised methods\footnote{\url{https://github.com/yulequan/UA-MT} \\ \url{https://github.com/kleinzcy/SASSnet} \\
\url{https://github.com/HiLab-git/DTC} \\ 
\url{https://github.com/ycwu1997/MC-Net}}.
We adapted these implementations to our datasets\footnote{\url{https://github.com/PierreRouge}}.
We also changed change the VNet \cite{milletari2016v} backbone architecture by a standard U-Net to have a fair comparison with the state of the art U-Net \cite{ronneberger2015u} supervised baseline.
All methods were trained with the same learning rate ($0.01$) and the same final weight $\lambda_{c}$ for the consistency loss ($0.01$).
For the supervised loss on labeled data, we used the combination of soft Dice loss and cross-entropy whereas for the consistency loss, we used the soft Dice loss only.

\subsection{Results}

\begin{figure*}[t]
    \centering
    \includegraphics[width=\linewidth]{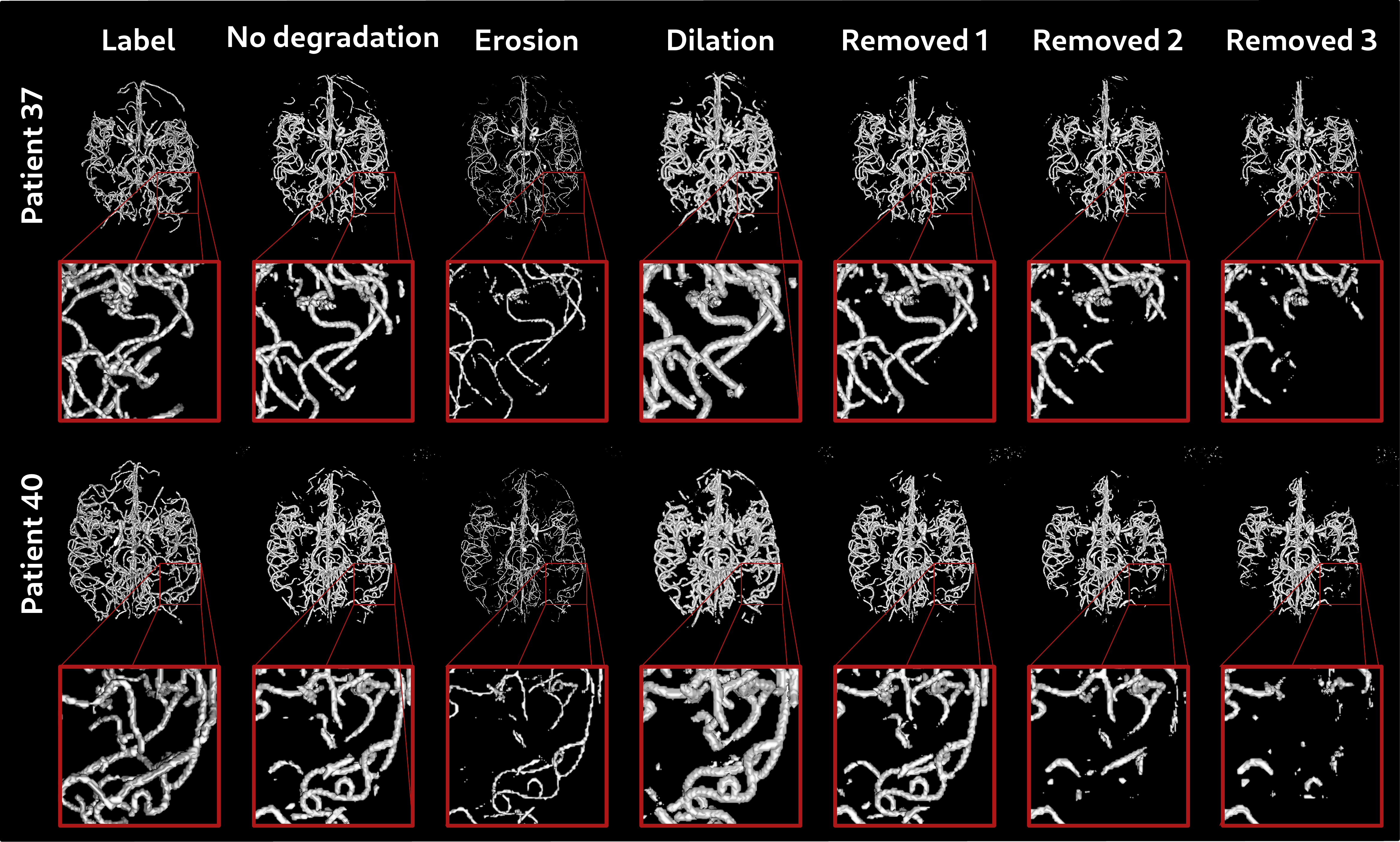}
    \caption{Visual results for two different patient from experiment 3. These segmentations are produced by the UA-MT model trained on the different damaged dataset (each column) with five training samples. Red boxes focus on some part of the cerebrovascular network to show better the differences between experiments.}
    \label{fig:visual_results_xp3}
\end{figure*}

\subsubsection{Experiment 1: Methods comparison and impact of the quantity of annotated data}

The results of these experiments are presented on Fig.~\ref{fig:res_xp1_bullitt} for the Bullitt dataset and Fig.~\ref{fig:res_xp1_IXI} for the IXI dataset. Firstly, we can see that all semi-supervised methods perform better than the supervised U-Net baseline, regardless of the number of labeled sample in the training set.  
However, the degree of improvement diminishes with the addition of more labeled samples, as expected.
For example the improvement in the $DSC$ between UA-MT and U-Net when using 1 labeled sample is $0.04$ but it decreases to $0.009$ when using 9 labeled samples on IXI dataset. Qualitative results confirm these findings (see Figs. \ref{fig:visual_results_xp1_Bullitt} and \ref{fig:visual_results_xp1_IXI}). In addition, we can observe that the segmentations generated by the supervised U-Net exhibit significant noise, as confirmed by the higher standard deviations, particularly when only one sample is used for training. This noise is typically attributed to an overfitting problem.


Another noteworthy result is the absence of a significant difference observed among the semi-supervised methods. Surprisingly, all methods exhibit similar performances irrespective of the number of labeled samples and for both datasets. This contrasts with findings in the literature, where more recent methods were reported to perform better on the LA dataset \cite{xiong2021global}. We assume that the higher complexity of the cerebrovascular structures and imperfections in the labeled data may be the cause of these divergent conclusions. Conducting benchmarks on other types of datasets is warranted to draw more comprehensive and generalizable conclusions.


\subsubsection{Experiment 2: Dependency to annotated data}

The results of this experiment are presented in Fig.~\ref{fig:seed_impact_Bullitt} and \ref{fig:seed_impact_IXI}. For each data composition (\textit{i.e.} number of labeled samples), we illustrated the segmentation results statistics over $9$ trainings with different samples.
This experiment confirm that the supervised baseline is less robust to the data variability than the semi-supervised approach,  as indicated by the significantly lower standard deviation on both metrics, regardless of the number of labeled samples.
This shows that the unsupervised regularization provided by the semi-supervised methods is widely effective to improve the generalization of the model. 

Furthermore, while the addition of more labeled data contributes to regularization, intriguingly, the utilization of a substantial amount of unlabeled data appears to be more effective in this regard. Therefore, semi-supervised methods can serve to enhance performance to some extent, but their primary value lies in effectively regularizing the segmentation model.

\subsubsection{Experiment 3: Impact of the annotation quality}

The results of this experiment are presented on Fig.~\ref{fig:deterioration_results}. 
The results show that training a model with annotations with a systematic under or over estimation of the vessel radius (dilation, removed 1/2/3) yield significantly worse results than with precise labels (reference). 
When only one labeled sample is used during training, the mean $DSC$ on the test set drops from $0.63$ (with the reference labels) to $0.52$ and $0.48$ with the dilated and eroded labels, respectively. This phenomenon persists regardless of the number of labeled samples in the training set. Moreover, the drop in $DSC$ slightly increases with the addition of more labeled samples, indicating that the regularization typically observed when adding data does not compensate for the error in the labels. This was expected, as the labels present a systematic error in the experiment.
Interestingly, the results obtained with the dilation experiment exhibit a clDice very similar to that of the reference experiment. Indeed, the clDice is not a suitable metric to assess over-segmentated results, as it compares the overlap of a centerline versus a segmentation. On the contrary, the results obtained with the erosion experiment exhibit a larger drop in $clDice$ than in $Dice$. When removing the volumic bias of the $Dice$ (\textit{i.e.} where larger vessels weight more than thinner vessels), the results are even worse than expected. This indicates that many vessels are missing in the segmentation, which is confirmed by visual inspection (see Fig~\ref{fig:visual_results_xp3}).

A compelling result from this experiment is that the segmentation from the removed 1/2/3 experiments are much better than that of dilation and erosion. 
This tends to demonstrates that when labeling vascular datasets, it is more important to accurately delineating the vessel borders than extensively annotating all vessels. 

%
Indeed, we can see in Fig.~\ref{fig:deterioration_results} that when just some distal vessels are missed (experiment removed 1) the results are not significantly degraded. They are even improved for $nsamples=1$, which again highlights the high sensibility of the results to the specific volume used for training (see experiment 2).
In experiments removed 2 and removed 3, there is a significant degradation of the results, but much less important than for the other types of degradations. Moreover, contrary to the dilation and erosion experiments, the $DSC$ increases regularly with the number of labeled samples. For $nsamples=2$, the drop of $DSC$ compared with the reference experiment is $0.08$ and $0.12$ for removed 2 and removed 3 respectively, and for $nsamples=5$, the drop decreases to $0.05$ and $0.09$. The same observation can be seen for the $clDice$. 
This shows a regularization effect when more labeled samples are added to the training set. The lack of some vessels in the labels introduces a noise in the annotations that is partially compensated when adding more partially annotated volumes. However, a  performance gap still subsists because the missing vessels in the annotations introduce a concept-shift (i.e. small vessels located at the distal part of the vascular network are labeled ``non vessels" in the training set but are labeled ``vessels" in the test set) which is not compensated by the increasing number of data. This highlights the problem of concept shift that occurs when different experts have different understanding of which vessels must be segmented.


\section{Discussion and conclusion}
\label{sec:discussion_conclusion}

In this study, we explored various semi-supervised methods across different data scenarios, aiming to discern the impact of both quantity and quality of annotations in the context of cerebrovascular segmentation. Our primary objective was to establish guidelines that optimize the annotation process and assist the community in making informed choices regarding model training. We address challenges such as annotation scarcity, noisy labels, and concept shift, particularly focusing on two common concept shift situations encountered in cerebrovascular segmentation: ambiguity at the vessel borders and uncertainty about which vessels should be included in the annotations.

Initially, our experiments indicate a significant advantage of using semi-supervised methods over a supervised model, especially when dealing with a low number of labeled samples. The extent of the benefit increases as the quantity of labeled samples decreases.
However, we showed that the performance gap between semi-supervised methods and the supervised baseline diminishes rapidly as the amount of annotated data increases. While this finding may temper expectations regarding the potential of semi-supervised methods to significantly enhance model performance, our experiments demonstrate their effectiveness in model regularization and reduction of overfitting. In the context of data scarcity, we believe that the most important added value of semi-supervised methods for cerebrovascular segmentation lies in these properties.

We shed light on the high impact of the concept shift in cerebrovascular segmentation. 
Such type of imperfection in the labeled data is very frequent in this application and appear when the annotation guidelines are not precise enough or when several experts perform the annotations.
Contrary to the presence of noise in the labels that could be regularized by adding more labeled data, a concept shift problem, such as the under or over estimation of the vessel border, drastically decreases the performance of a model. 

Based on our experiments, we can formulate several guidelines both regarding the annotation process of cerebrovascular images and the training of segmentation models.
Firstly, when training a model using an non in-house labeled dataset, it is very important to carefully analyze it to understand the chosen concept of a vessel during the annotation process. If the model exhibits poor generalization to a test dataset, the cause may be attributed to the concept shift between the training and test datasets.
Secondly, when creating a new labeled dataset, it is of the utmost importance to define a precise policy to reduce the concept shift in the dataset. In particular, the following questions should be addressed : which vessels should be includes in the annotation? when to stop labeling a vessel? what is the degree of precision of the vessel border that is expected.
Finally, the label quality is more important than their quantity. When there is a given time slot of expert availability, the primary focus should be on creating precise annotations, rather than annotating more samples with lower quality. In particular, precision in delineating vessel borders is more important than labeling extensively all the vessels up to the image resolution limit.

\section*{Acknowledgements}
This work was supported by the French \emph{Agence Nationale de la Recherche} (grant ANR-20-CE45-0011) as part of the PreSPIN project. This work was also partly funded by France Life Imaging (grant ANR-11-INBS-0006). This work was granted access to the HPC resources of IDRIS under the allocation 2023-AD011013610 made by GENCI.



 \bibliographystyle{elsarticle-num} 
 \bibliography{biblio}

\begin{thebibliography}{10}
\expandafter\ifx\csname url\endcsname\relax
  \def\url#1{\texttt{#1}}\fi
\expandafter\ifx\csname urlprefix\endcsname\relax\def\urlprefix{URL }\fi
\expandafter\ifx\csname href\endcsname\relax
  \def\href#1#2{#2} \def\path#1{#1}\fi

\bibitem{conze2023current}
P.-H. Conze, G.~Andrade-Miranda, V.~K. Singh, V.~Jaouen, D.~Visvikis, Current and emerging trends in medical image segmentation with deep learning, IEEE Transactions on Radiation and Plasma Medical Sciences (2023).

\bibitem{ronneberger2015u}
O.~Ronneberger, P.~Fischer, T.~Brox, {U-Net}: {C}onvolutional networks for biomedical image segmentation, in: Medical Image Computing and Computer-Assisted Intervention (MICCAI), Proceedings, Part III, 2015, pp. 234--241.

\bibitem{isensee2021nnu}
F.~Isensee, P.~F. Jaeger, S.~A.~A. Kohl, J.~Petersen, K.~H. Maier-Hein, {nnU-Net}: {A} self-configuring method for deep learning-based biomedical image segmentation, Nature Methods 18~(2) (2021) 203--211.

\bibitem{azad2022medical}
R.~Azad, E.~K. Aghdam, A.~Rauland, Y.~Jia, A.~H. Avval, A.~Bozorgpour, S.~Karimijafarbigloo, J.~P. Cohen, E.~Adeli, D.~Merhof, Medical image segmentation review: The success of {U-Net}, arXiv preprint arXiv:2211.14830 (2022).

\bibitem{chen2021transunet}
J.~Chen, Y.~Lu, Q.~Yu, X.~Luo, E.~Adeli, Y.~Wang, L.~Lu, A.~L. Yuille, Y.~Zhou, {TransUNet}: Transformers make strong encoders for medical image segmentation, arXiv preprint arXiv:2102.04306 (2021).

\bibitem{hatamizadeh2021swin}
A.~Hatamizadeh, V.~Nath, Y.~Tang, D.~Yang, H.~R. Roth, D.~Xu, Swin {UNETR}: Swin transformers for semantic segmentation of brain tumors in {MRI} images, in: International MICCAI Brainlesion Workshop, 2021, pp. 272--284.

\bibitem{shamshad2023transformers}
F.~Shamshad, S.~Khan, S.~W. Zamir, M.~H. Khan, M.~Hayat, F.~S. Khan, H.~Fu, Transformers in medical imaging: A survey, Medical Image Analysis 88 (2023) 102802.

\bibitem{DBLP:journals/mia/LesageABF09}
D.~Lesage, E.~D. Angelini, I.~Bloch, G.~Funka{-}Lea, A review of {3D} vessel lumen segmentation techniques: {M}odels, features and extraction schemes, Medical Image Analysis 13 (2009) 819--845.

\bibitem{DBLP:journals/cmpb/MocciaMHM18}
S.~Moccia, E.~D. Momi, S.~E. Hadji, L.~S. Mattos, Blood vessel segmentation algorithms - {R}eview of methods, datasets and evaluation metrics, Computer Methods and Programs in Biomedicine 158 (2018) 71--91.

\bibitem{sanchesa2019cerebrovascular}
P.~Sanches, C.~Meyer, V.~Vigon, B.~Naegel, Cerebrovascular network segmentation of {MRA} images with deep learning, in: IEEE International Symposium on Biomedical Imaging (ISBI), Proceedings, 2019, pp. 768--771.

\bibitem{livne2019u}
M.~Livne, J.~Rieger, O.~U. Aydin, A.~A. Taha, E.~M. Akay, T.~Kossen, J.~Sobesky, J.~D. Kelleher, K.~Hildebrand, D.~Frey, V.~I. Madai, A {U-Net} deep learning framework for high performance vessel segmentation in patients with cerebrovascular disease, Frontiers in Neuroscience 13 (2019) 97.

\bibitem{Valderrama2023}
N.~Valderrama, I.~Pitsiorlas, L.~Vargas, P.~Arbelaez, M.~A. Zuluaga, {JOB-VS}: {J}oint brain-vessel segmentation in {TOF-MRA} images, in: IEEE International Symposium on Biomedical Imaging (ISBI), Proceedings, 2023.

\bibitem{lin2023high}
F.~Lin, Y.~Xia, S.~Song, N.~Ravikumar, A.~F. Frangi, High-throughput {3DRA} segmentation of brain vasculature and aneurysms using deep learning, Computer Methods and Programs in Biomedicine 230 (2023) 107355.

\bibitem{mou2021cs2}
L.~Mou, Y.~Zhao, H.~Fu, Y.~Liu, J.~Cheng, Y.~Zheng, P.~Su, J.~Yang, L.~Chen, A.~F. Frangi, M.~Akiba, J.~Liu, {CS2-Net}: {D}eep learning segmentation of curvilinear structures in medical imaging, Medical Image Analysis 67 (2021) 101874.

\bibitem{zhang2020cerebrovascular}
H.~Zhang, L.~Xia, R.~Song, J.~Yang, H.~Hao, J.~Liu, Y.~Zhao, Cerebrovascular segmentation in {MRA} via reverse edge attention network, in: Medical Image Computing and Computer Assisted Intervention (MICCAI), Proceedings, Part VI, 2020, pp. 66--75.

\bibitem{ciecholewski2021computational}
M.~Ciecholewski, M.~Kassja{\'n}ski, Computational methods for liver vessel segmentation in medical imaging: A review, Sensors 21~(6) (2021) 2027.

\bibitem{yang2023benchmarking}
K.~Yang, F.~Musio, Y.~Ma, N.~Juchler, J.~C. Paetzold, R.~Al-Maskari, L.~Höher, H.~B. Li, I.~E. Hamamci, A.~Sekuboyina, S.~Shit, H.~Huang, D.~Waldmannstetter, F.~Kofler, F.~Navarro, M.~Menten, I.~Ezhov, D.~Rueckert, I.~Vos, Y.~Ruigrok, B.~Velthuis, H.~Kuijf, J.~Hämmerli, C.~Wurster, P.~Bijlenga, L.~Westphal, J.~Bisschop, E.~Colombo, H.~Baazaoui, A.~Makmur, J.~Hallinan, B.~Wiestler, J.~S. Kirschke, R.~Wiest, E.~Montagnon, L.~Letourneau-Guillon, A.~Galdran, F.~Galati, D.~Falcetta, M.~A. Zuluaga, C.~Lin, H.~Zhao, Z.~Zhang, S.~Ra, J.~Hwang, H.~Park, J.~Chen, M.~Wodzinski, H.~Müller, P.~Shi, W.~Liu, T.~Ma, C.~Yalçin, R.~E. Hamadache, J.~Salvi, X.~Llado, U.~M. L.-T. Estrada, V.~Abramova, L.~Giancardo, A.~Oliver, J.~Liu, H.~Huang, Y.~Cui, Z.~Lin, Y.~Liu, S.~Zhu, T.~R. Patel, V.~M. Tutino, M.~Orouskhani, H.~Wang, M.~Mossa-Basha, C.~Zhu, M.~R. Rokuss, Y.~Kirchhoff, N.~Disch, J.~Holzschuh, F.~Isensee, K.~Maier-Hein, Y.~Sato, S.~Hirsch, S.~Wegener, B.~Menze, Benchmarking the cow with the topcow challenge:
  Topology-aware anatomical segmentation of the circle of willis for cta and mra (2023).

\bibitem{aylward2002initialization}
S.~R. Aylward, E.~Bullitt, Initialization, noise, singularities, and scale in height ridge traversal for tubular object centerline extraction, IEEE Transactions on Medical Imaging 21~(2) (2002) 61--75.

\bibitem{ixidataset}
\url{https://brain-development.org/ixi-dataset/}.

\bibitem{liu2022deep}
X.~Liu, C.~Yoo, F.~Xing, H.~Oh, G.~El~Fakhri, J.-W. Kang, J.~Woo, Deep unsupervised domain adaptation: A review of recent advances and perspectives, APSIPA Transactions on Signal and Information Processing 11~(1) (2022).

\bibitem{kouw2018introduction}
W.~M. Kouw, M.~Loog, An introduction to domain adaptation and transfer learning, arXiv preprint arXiv:1812.11806 (2018).

\bibitem{jiao2022learning}
R.~Jiao, Y.~Zhang, L.~Ding, R.~Cai, J.~Zhang, Learning with limited annotations: {A} survey on deep semi-supervised learning for medical image segmentation, arXiv preprint arXiv:2207.14191 (2022).

\bibitem{shi2021inconsistency}
Y.~Shi, J.~Zhang, T.~Ling, J.~Lu, Y.~Zheng, Q.~Yu, L.~Qi, Y.~Gao, Inconsistency-aware uncertainty estimation for semi-supervised medical image segmentation, IEEE Transactions on Medical Imaging 41~(3) (2021) 608--620.

\bibitem{wang2022ssa}
X.~Wang, Y.~Yuan, D.~Guo, X.~Huang, Y.~Cui, M.~Xia, Z.~Wang, C.~Bai, S.~Chen, {SSA-Net}: {S}patial self-attention network for {COVID-19} pneumonia infection segmentation with semi-supervised few-shot learning, Medical Image Analysis 79 (2022) 102459.

\bibitem{thompson2022pseudo}
B.~H. Thompson, G.~Di~Caterina, J.~P. Voisey, Pseudo-label refinement using superpixels for semi-supervised brain tumour segmentation, in: IEEE International Symposium on Biomedical Imaging (ISBI), Proceedings, 2022, pp. 1--5.

\bibitem{you2022simcvd}
C.~You, Y.~Zhou, R.~Zhao, L.~Staib, J.~S. Duncan, {SimCVD}: Simple contrastive voxel-wise representation distillation for semi-supervised medical image segmentation, IEEE Transactions on Medical Imaging 41~(9) (2022) 2228--2237.

\bibitem{zheng2019semi}
H.~Zheng, L.~Lin, H.~Hu, Q.~Zhang, Q.~Chen, Y.~Iwamoto, X.~Han, Y.-W. Chen, R.~Tong, J.~Wu, Semi-supervised segmentation of liver using adversarial learning with deep atlas prior, in: Medical Image Computing and Computer Assisted Intervention (MICCAI), Proceedings, Part VI, 2019, pp. 148--156.

\bibitem{tarvainen2017mean}
A.~Tarvainen, H.~Valpola, Mean teachers are better role models: {W}eight-averaged consistency targets improve semi-supervised deep learning results, in: International Conference on Neural Information Processing Systems (NeurIPS), Proceedings, 2017, pp. 1195--1204.

\bibitem{luo2021semi}
X.~Luo, J.~Chen, T.~Song, G.~Wang, Semi-supervised medical image segmentation through dual-task consistency, in: AAAI Conference on Artificial Intelligence (AAAI), Proceedings, 2021, pp. 8801--8809.

\bibitem{yu2019uncertainty}
L.~Yu, S.~Wang, X.~Li, C.-W. Fu, P.-A. Heng, Uncertainty-aware self-ensembling model for semi-supervised {3D} left atrium segmentation, in: Medical Image Computing and Computer Assisted Intervention (MICCAI), Proceedings, Part II, 2019, pp. 605--613.

\bibitem{wu2021semi}
Y.~Wu, M.~Xu, Z.~Ge, J.~Cai, L.~Zhang, Semi-supervised left atrium segmentation with mutual consistency training, in: Medical Image Computing and Computer Assisted Intervention (MICCAI), Proceedings, Part II, 2021, pp. 297--306.

\bibitem{lei2022semi}
T.~Lei, D.~Zhang, X.~Du, X.~Wang, Y.~Wan, A.~K. Nandi, Semi-supervised medical image segmentation using adversarial consistency learning and dynamic convolution network, IEEE Transactions on Medical Imaging (2022).

\bibitem{bai2023bidirectional}
Y.~Bai, D.~Chen, Q.~Li, W.~Shen, Y.~Wang, Bidirectional copy-paste for semi-supervised medical image segmentation, in: Proceedings of the IEEE/CVF Conference on Computer Vision and Pattern Recognition, 2023, pp. 11514--11524.

\bibitem{luo2022semi}
X.~Luo, M.~Hu, T.~Song, G.~Wang, S.~Zhang, Semi-supervised medical image segmentation via cross teaching between {CNN} and transformer, in: International Conference on Medical Imaging with Deep Learning (MIDL), Proceedings, 2022, pp. 820--833.

\bibitem{xia20203d}
Y.~Xia, F.~Liu, D.~Yang, J.~Cai, L.~Yu, Z.~Zhu, D.~Xu, A.~Yuille, H.~Roth, {3D} semi-supervised learning with uncertainty-aware multi-view co-training, in: IEEE/CVF Winter Conference on Applications of Computer Vision (WACV), Proceedings, 2020, pp. 3646--3655.

\bibitem{li2020shape}
S.~Li, C.~Zhang, X.~He, Shape-aware semi-supervised {3D} semantic segmentation for medical images, in: Medical Image Computing and Computer Assisted Intervention (MICCAI), Proceedings, Part I, 2020, pp. 552--561.

\bibitem{zhang2017deep}
Y.~Zhang, L.~Yang, J.~Chen, M.~Fredericksen, D.~P. Hughes, D.~Z. Chen, Deep adversarial networks for biomedical image segmentation utilizing unannotated images, in: Medical Image Computing and Computer Assisted Intervention (MICCAI), Proceedings, Part III, 2017, pp. 408--416.

\bibitem{hou2022semi}
J.~Hou, X.~Ding, J.~D. Deng, Semi-supervised semantic segmentation of vessel images using leaking perturbations, in: IEEE/CVF Winter Conference on Applications of Computer Vision (WACV), Proceedings, 2022, pp. 2625--2634.

\bibitem{hang2020local}
W.~Hang, W.~Feng, S.~Liang, L.~Yu, Q.~Wang, K.-S. Choi, J.~Qin, Local and global structure-aware entropy regularized mean teacher model for {3D} left atrium segmentation, in: Medical Image Computing and Computer Assisted Intervention (MICCAI), Proceedings, Part I, 2020, pp. 562--571.

\bibitem{xie2022semi}
L.~Xie, Z.~Chen, X.~Sheng, Q.~Zeng, J.~Huang, C.~Wen, L.~Wen, G.~Xie, Y.~Feng, Semi-supervised region-connectivity-based cerebrovascular segmentation for time-of-flight magnetic resonance angiography image, Computers in Biology and Medicine 149 (2022) 105972.

\bibitem{chen2022generative}
C.~Chen, K.~Zhou, Z.~Wang, R.~Xiao, Generative consistency for semi-supervised cerebrovascular segmentation from {TOF-MRA}, IEEE Transactions on Medical Imaging 42~(2) (2022) 346--353.

\bibitem{dice1945measures}
L.~R. Dice, Measures of the amount of ecologic association between species, Ecology 26 (1945) 297--302.

\bibitem{shit2021cldice}
S.~Shit, J.~C. Paetzold, A.~Sekuboyina, I.~Ezhov, A.~Unger, A.~Zhylka, J.~P.~W. Pluim, U.~Bauer, B.~H. Menze, {clDice}--{A} novel topology-preserving loss function for tubular structure segmentation, in: IEEE/CVF Conference on Computer Vision and Pattern Recognition (CVPR), Proceedings, 2021, pp. 16560--16569.

\bibitem{milletari2016v}
F.~Milletari, N.~Navab, S.-A. Ahmadi, V-net: Fully convolutional neural networks for volumetric medical image segmentation, in: 2016 fourth international conference on 3D vision (3DV), Ieee, 2016, pp. 565--571.

\bibitem{xiong2021global}
Z.~Xiong, Q.~Xia, Z.~Hu, N.~Huang, C.~Bian, Y.~Zheng, S.~Vesal, N.~Ravikumar, A.~Maier, X.~Yang, et~al., A global benchmark of algorithms for segmenting the left atrium from late gadolinium-enhanced cardiac magnetic resonance imaging, Medical image analysis 67 (2021) 101832.

\end{thebibliography}

\end{document}